\documentclass[aps,pra,floats,floatfix,twocolumn,longbibliography]{revtex4-1}

\usepackage{graphicx}
\usepackage{epstopdf}
\graphicspath{{./Figs/}{./figs/}}

\usepackage{latexsym}
\usepackage{amsmath}
\usepackage{amssymb}
\usepackage{bm}
\usepackage{wasysym}

\usepackage{color}
\usepackage[colorlinks,allcolors=blue,bookmarks=true]{hyperref}
\usepackage{flafter}
\usepackage[percent]{overpic}

\newcommand{\eexp}[1]{\mathrm{e}^{#1}}

\newcommand{\ket}[1]{\left| #1 \right\rangle}
\newcommand{\braket}[1]{\left\langle #1 \right\rangle}

\newcommand{\ora}{\protect\overrightarrow}

\newcommand{\be}[1]{\begin{eqnarray}{\label{e#1}}} 
\newcommand{\beq}{\begin{eqnarray}}
\newcommand{\eeq}{\end{eqnarray}} 

\newcommand{\hide}[1]{}
\newcommand{\Eq}[1]{\textcolor{blue}{{Eq.}\!\!~(\ref{#1})}}
\newcommand{\App}[1]{\textcolor{blue}{{Appendix}\!~\ref{#1}}}

\newcommand{\Fig}[1]{\textcolor{blue}{Fig.}\!\!~\ref{#1}}

\definecolor{myred}{rgb}  {0.5,0.0,0.0}
\newcommand{\rmrk}[1] {#1} 

\newcommand{\sect}[1]{{\bf #1.--}}

\begin{document}

\title{Monodromy and chaos for condensed bosons in optical lattices}

\author{Geva Arwas$^{1,2}$, Doron Cohen$^1$}

\affiliation{
\mbox{$^1$Department of Physics, Ben-Gurion University of the Negev, Beer-Sheva 84105, Israel} \\
\mbox{$^2$Department of Physics of Complex Systems, Weizmann Institute of Science, Rehovot 7610001, Israel}
}

\begin{abstract}
We introduce a theory for the stability of a condensate in an optical lattice. We show that the understanding of the stability-to-ergodicity transition involves the fusion of {\em monodromy} and {\em chaos} theory. Specifically, the condensate can decay if a connected chaotic pathway to depletion is formed, which requires swap of seperatrices in phase-space. 
\end{abstract}

\maketitle

\section{Introduction}

Ergodicity, as opposed to Stability, is the threat that looms over the condensation of bosons in optical lattices. A major question of interest is whether an initial condensate is likely to be depleted. The simplest setup is the dimer, aka Bosonic Josephson Junction \cite{maya,PhysRevLett.95.010402,levy2007ac}, where condensation in the upper orbital can become unstable if the interaction exceeds a critical value. A more challenging setup is a ring lattice \cite{Amico2014,sfa,Paraoanu,Hallwood06,gallemi2015fragmented}, where the particles are condensed into an excited momentum orbital. If such flow-state is metastable, it can be regarded as a mesoscopic version of supefluidity. It has been realized that the theory for this superfluidity requires analysis that goes beyond the tradition framework of Landau and Bogoliubov, because the underlying dynamics is largely chaotic \cite{KolovskyReview,sfc}. 

The structure of the classical phase-space is reflected in the quantum spectrum, and provides the key for quantum-chaos theory of mesoscopic superfluity. 
In the present work we highlight the essential ingredient for the crossover from stability to ergodicity. We consider the minimal setup: a 3 site ring. We show that the understanding of this transition involves the fusion of two major research themes: {\em monodromy} and {\em chaos}. 
   
\sect{Monodromy}
%
%
The dynamics of an integrable (non-chaotic) system, for a given value of the conserved constants-of-motion, can be described by a set of action-angle variables, that parametrize a torus in phase space. In systems with monodromy, they cannot be defined globally:
due to the non trivial topology of phase space, the action-angle variables cannot be identified in a continuous way in the parameter-space that is formed by the conserved quantities \cite{Duistermaat,monodromybook}. 
Accordingly, it is impossible to describe the quantum spectrum by a global set of good quantum numbers  \cite{cushman1988quantum,vungoc}.
Rather, the good quantum numbers (quantized ``actions") that are implied by the EBK quantization scheme
form a lattice that features a topological defect \cite{Zhilinskii2006}. 
Such Hamiltonian monodromy is found in many physical systems, such as the spherical \cite{cushman1988quantum,PhysRevA.69.032504} and the swing-spring  \cite{PhysRevLett.93.024302,PhysRevLett.103.034301} pendula, Spin-1 condensed bosons \cite{PhysRevA.83.033605}, the Dicke model \cite{monodromydicke}, and even the hydrogen atom \cite{monodromyhydrogen}. 
A dynamical manifestation of monodromy in a classical system has been recently demonstrated \cite{dynamicalmonodromyexp}.

\sect{Chaos}
The condensation of particles in a single orbital is a many-body coherent state. It can be represented in phase-space as a Gaussian-like distribution that is supported by a stationary point (SP). If this SP is the minimum of the energy landscape, it is known as Landau energetic stability, and leads, for a clean ring, to the Landau criterion for superfluidity. More generally one has to find the Bogoliubov excitations $\omega_r$ of condensate. If some of the frequencies become complex, the SP is considered to be unstable. What we have demonstrated in previous work \cite{sfc,sfa} was that this type of local stability analysis does not provide the required criteria for stability. Rather, in order to determine whether the system will ergodize, it is essential to study the  global structure of phase-space, and to take into account the role of chaos.

\sect{Connectivity}
The major insight can be described schematically as follows. Let us regard the SP that supports the condensate as the {\em origin} of phase-space. And let us regard the region that supports the completely depleted states as the {\em perimeter} of phase-space. The crucial question is whether there is a dynamical pathway that leads from the origin to the perimeter. We have observed numerically in \cite{sfc} that the formation of such pathway requires a swap of phase-space separatrices. But a theory for this swap transition has not been provided.

\sect{Outline}
\rmrk{For pedagogical purpose we first consider the stability question for the dimer. Then we go to the trimer and write its Hamiltonian} as the sum of integrable part $\mathcal{H}^{(0)}$, and additional terms $\mathcal{H}^{(\pm)}$ that induce the chaos. An example for the classical and quantum spectra is presented in \Fig{f1}. The spectra exhibit monodromy that we analyze in detail: the quantum monodromy is a reflection of the classical one. Then we explain how the spectrum is affected by changing a control parameter (detuning). In an hysteresis experiment \cite{NIST2} the detuning is determined by the rotation frequency of the device and the interaction strength between the bosons. We provide a geometrical explanation for the {\em swap transition}, and clarify the role of chaos in the de-stabilization of the condensate. 
\rmrk{In the summary section we point out the relevance of our study to the more general theme of thermalization in Bose-Hubbard chains}.

\section{The Model}

The Bose-Hubbard Hamiltonian (BHH) is a prototype model for cold atoms in optical lattices that has inspired state-of-the-art experiments \cite{BHH1,BHH2}, \rmrk{and has been proposed as a platform for quantum simulations. It describes a system of $N$ bosons in $L$ sites. The ring geometry in particular has attracted attention for atomtronic circuits \cite{Amico2014}.}
Taking into account that ${N}$ is constant of motion, the system has ${f=L{-}1}$ degrees of freedoms.  
The simplest configuration is the dimer ($L{=}2$), aka Bosonic Josephson Junction. 
Our main interest is in the minimal non-integrable configuration, which is the trimer ($L{=}3$). 
Below we briefly refer to the dimer Hamiltonian, and then turn to discuss the trimer Hamiltonian. 
Further technical details about the latter are provided in \App{sA}.

\subsection{The dimer}

\rmrk{The Hamiltonian of the dimer is
\be{A0} 
\mathcal{H}_{\text{dimer}} = 
- \frac{K}{2} \left({a}_{2}^{\dag} {a}_{1} + {a}_{1}^{\dag} {a}_{2}\right)
+ \frac{U}{2} \sum_{j=1}^{2} {a}_{j}^{\dag}  {a}_{j}^{\dag}  {a}_{j} {a}_{j} 
\ \ \ \ \ 
\eeq}
where $K$ is the hopping frequency, $U$ is the on-site interaction, 
The $ {a}_{j}$ and $ {a}_{j}^{\dag}$ are the Bosonic annihilation and creation operators.
The total number of particles $N=\sum_j {a}_{j}^{\dag} {a}_{j}$ is a constant of motion.

\rmrk{It is convenient to switch to orbital representation.
One defines annihilation and creation operators ${b}_{k}$ and ${b}_{k}^{\dag}$,  
such that $b_{\pm}^{\dag} = \frac{1}{ \sqrt{2} } (a_1^{\dag} \pm a_2^{\dag})$ 
creates bosons in the lower and upper orbitals.
For the purpose of semiclassical treatment we define action-angle variables via 
\beq
b_k \ \ = \ \ \sqrt{n_k} \ e^{i\varphi_k}
\eeq
After dropping an $N$ dependent constant the Hamiltonian takes the form
\be{A3}
\mathcal{H}_{\text{dimer}}(\tilde{\varphi},\tilde{n}) \ = \  
- \mathcal{E} \tilde{n} + 
\frac{U}{2} (N{-}\tilde{n}) \, \tilde{n} \, [1+\cos(2\tilde{\varphi})] \ \
\eeq
where $\tilde{n}=n_{+}$ is the occupation of the~($+$) orbital,
and ${\mathcal{E} = K}$ is the detuning (energy difference between the orbitals).
The angle ${\tilde{\varphi} = (\varphi_{+}-\varphi_{-})}$ 
serves as a conjugate coordinate.   
The phase space of this Hamiltonian has the topology of Bloch sphere.
The Hamiltonian possesses two SPs that are located at ${\tilde{n}{=}0}$ and ${\tilde{n}{=}N}$, 
which are the North pole and the South poles of the Bloch sphere. }

\subsection{The trimer}

The BHH for $L$ sites in a ring geometry is 
\be{A1} 
\mathcal{H} = \sum_{j=1}^{L} \left[
\frac{U}{2}  {a}_{j}^{\dag}  {a}_{j}^{\dag}  {a}_{j} {a}_{j} 
- \frac{K}{2} \left(\eexp{i(\Phi/L)}  {a}_{j{+}1}^{\dag} {a}_{j} + \text{h.c.} \right)
\right] \ \ \ \ \ 
\eeq
where $j$ mod$(L)$ labels the sites of the ring, 
and other notations are as in the dimer case. 
The so-called Sagnac phase $\Phi$ is proportional to 
the rotation frequency of the device: it can be 
regarded as the Aharonov-Bohm flux that is associated
with Coriolis field in the rotating frame~\cite{fetter,NIST2}.

\rmrk{It is convenient to switch to momentum representation.  
One defines annihilation and creation operators ${b}_{k}$ and ${b}_{k}^{\dag}$,  
such that $b_k^{\dag} = \frac{1}{ \sqrt{L} } \sum_j \eexp{ikj} a_j^{\dag}$ 
creates bosons in the $k$-th momentum orbitals.}
Here we consider a 3-site ring that has 3~momentum orbitals labeled by their wavenumber~${k=0,1,2}$.  
Later we assume, without loss of generality, that the particles are initially condensed 
in the ${k=0}$ orbital.  This is not necessarily the ground-state orbital, 
because we allow the possibility that the ring is in a rotating frame.
After some time the condensate can be partially depleted
such that the occupation is ${(N{-}n_1{-}n_2,n_1,n_2)}$.

\rmrk{Since we have here an effectively 2~freedom system, it is convenient 
to define relative phases ${q_1 = \varphi_1 {-} \varphi_0 }$ and ${q_2 = \varphi_2 {-} \varphi_0 }$.}
Consequently the Hamiltonian can be written in terms of canonical coordinates as (\App{sA}):
\be{1}
\mathcal{H}(\varphi,n;\phi,M) = \mathcal{H}^{(0)}(\varphi,n;M) +  \left[\mathcal{H}^{(+)} + \mathcal{H}^{(-)} \right] \ \ 
\eeq
\rmrk{Here ${n=(n_1+n_2)/2}$ and ${M=(n_1-n_2)/2}$, 
while the conjugate angle variables are 
${\varphi = q_1 + q_2 }$ and ${\phi = q_1 - q_2  }$.}
The first term $\mathcal{H}^{(0)}$ is an integrable piece of the Hamiltonian that has~$M$ as an additional constant of motion: 
\be{2}
&& \mathcal{H}^{(0)}(\varphi,n;M) \ = \ \mathcal{E} n  + \mathcal{E}_{\perp} M - \frac{U}{3}M^2   \\ \nonumber
&& \ \ + \frac{2U}{3} (N-2n)\left[ \frac{3}{4}n + \sqrt{n^2-M^2} \cos(\varphi) \right] 
\eeq  
where $U$ is the interaction between the bosons, 
while the detuning ${\mathcal{E}}$ determines the energy difference between 
the condensate (${n=0}$) and the depleted states (${n=N/2}$).    
If we linearized $\mathcal{H}$ with respect to the ${(n_1,n_2)}$ occupations, 
we would get the Bogoliubov approximation, which is \Eq{e2}
without the third term ($M^2$), and with ${(N{-}2n)\approx N}$.
The additional terms $\mathcal{H}^{(\pm)}$ induce resonances that spoil the integrability, and give rise to chaos. 
\be{3}
\mathcal{H}^{(\pm)} = \frac{2U}{3} \sqrt{(N{-}2n) (n {\pm} M)} (n {\mp} M) \cos \left(\frac{3\phi {\mp} \varphi}{2} \right) \ \ \ \ 
\eeq  
\rmrk{Note that classically the total number of particles $N$ can be removed from the Hamiltonian by a simple scaling of $n$ and $M$. But upon quantization $1/N$ effectively plays the role of $\hbar$. It follows that the coherent state that is formed by condensation of the particles into a single orbital is represented in phase-space by a Gaussian-like distribution of uncertainty width $1/N$. See for example \cite{KolovskyReview,sfc}. The significance of this $1/N$ scale for the analysis of instabilities due to non-linear resonances has been illuminated in \cite{sfa}. }

\begin{figure}
\begin{center}
\begin{overpic}[width=0.95\hsize]{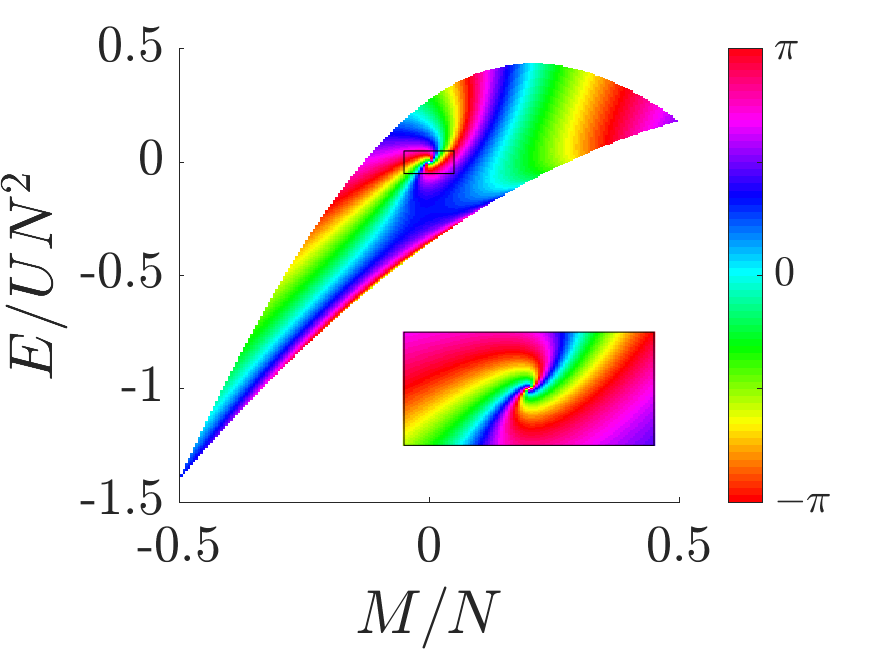} \put (25,66) {(a)} \end{overpic}
\begin{overpic}[width=0.95\hsize]{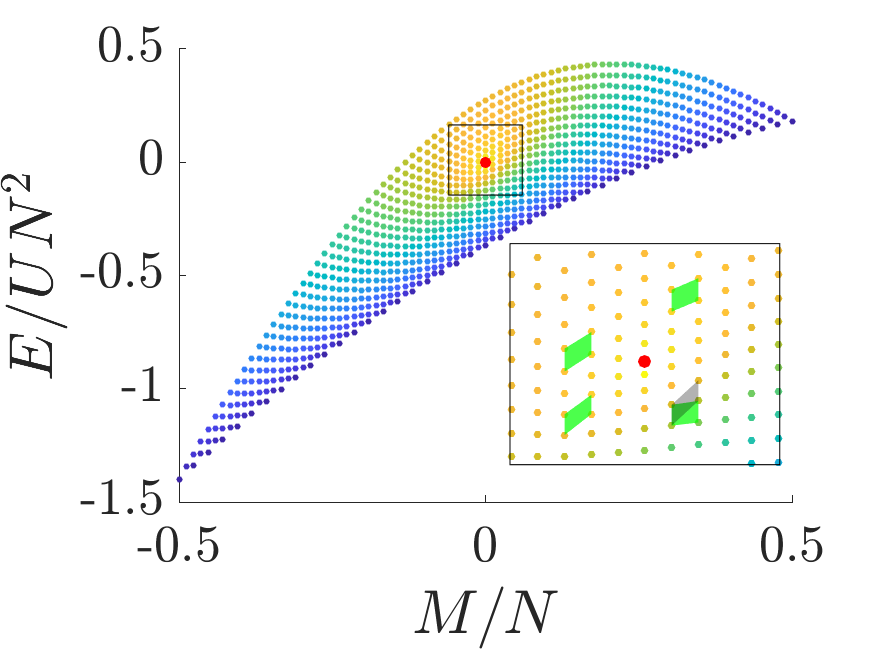} \put (25,64) {(b)} \end{overpic}
\caption{ \label{f1} 
{\bf Monodromy.} 
The classical and quantum spectra of the Hamiltonian $\mathcal{H}^{(0)}$. This Hamiltonian has a constant of motion~$M$, that describes the occupation imbalance of the ${k\neq 0}$ orbitals. In~(a) each point represents an $(M,E)$ torus in phase space, and the points are color coded by the value of a classical phase ($\beta$) that characterizes the torus. In~(b) each point represent an $\ket{M,E}$ eigenstate of $N=42$ particles, and the points are color coded by the expectation value of the variable~$n$, which is the total occupation of the ${k\neq 0}$ orbitals. Yellow color (${n < N/8}$) indicates a nearly coherent condensate, while blue implies a depleted eigenstate. In both panels ${\mathcal{E}/NU \approx -1/4}$ and ${\mathcal{E}_{\perp}/NU \approx 1/2}$.
The inset provides a zoom that demonstrates the monodromy: a topological defect in the lattice arrangement of the spectrum. 
\hfill
} 
\end{center}
\end{figure}

\begin{figure*}
\begin{center}
\begin{overpic}[trim={3.0cm 0 2.5cm 0},clip,width=0.132\hsize]{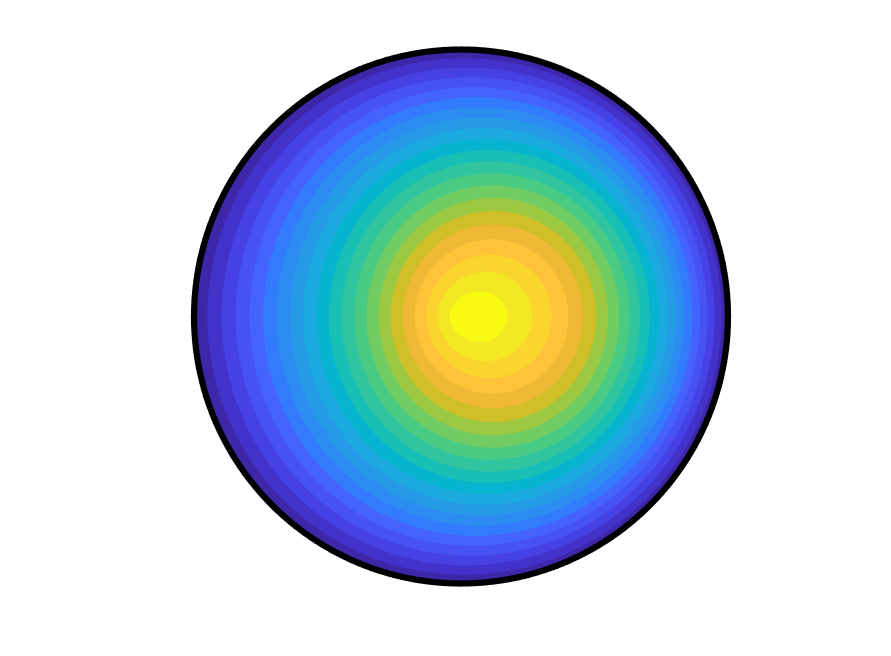}  \put (35,97) {(a)} \end{overpic}
\begin{overpic}[trim={3.0cm 0 2.5cm 0},clip,width=0.13\hsize]{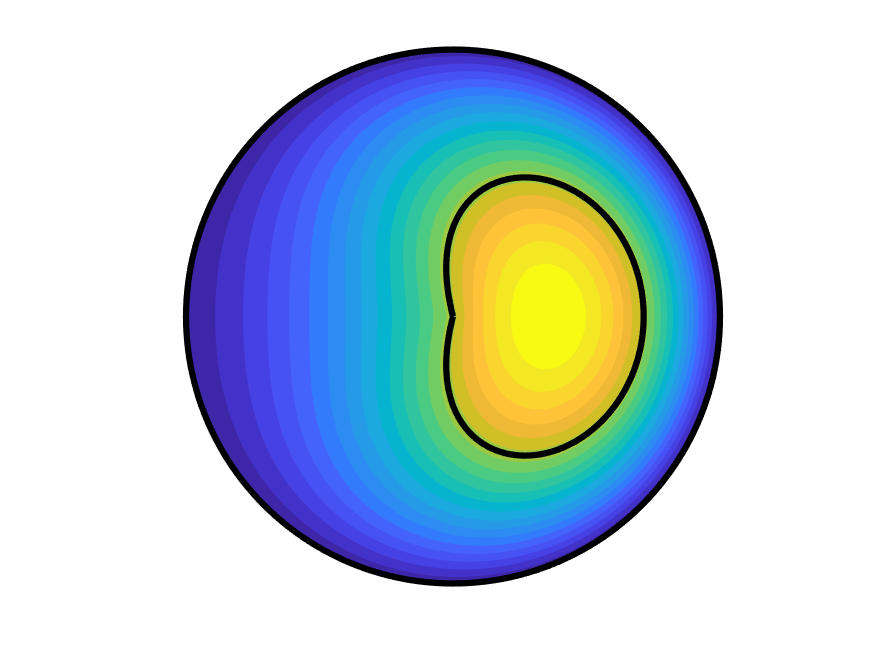}   \put (35,97) {(b)} \end{overpic}
\begin{overpic}[trim={3.0cm 0 2.5cm 0},clip,width=0.13\hsize]{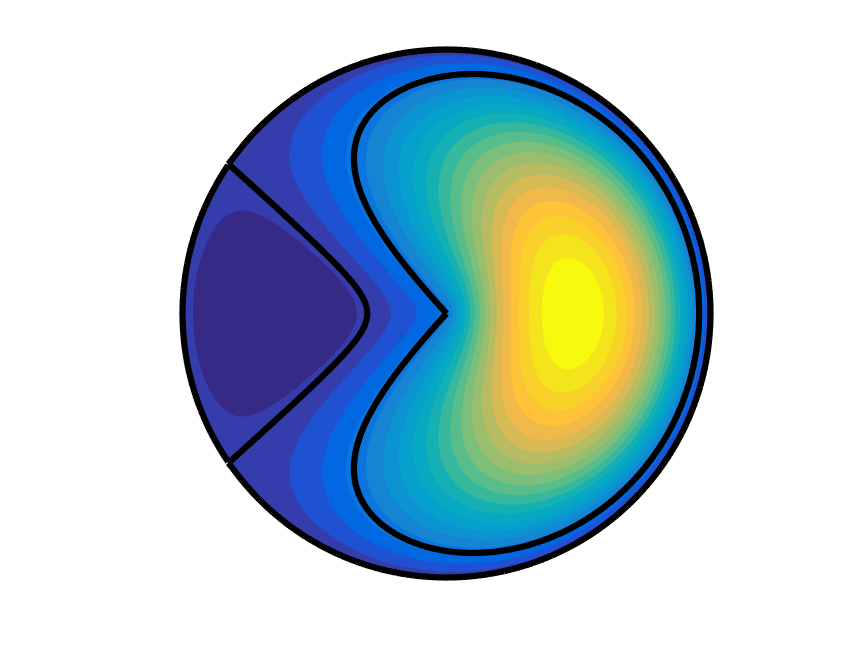}  \put (35,97) {(c)} \end{overpic}
\begin{overpic}[trim={3.0cm 0 2.5cm 0},clip,width=0.13\hsize]{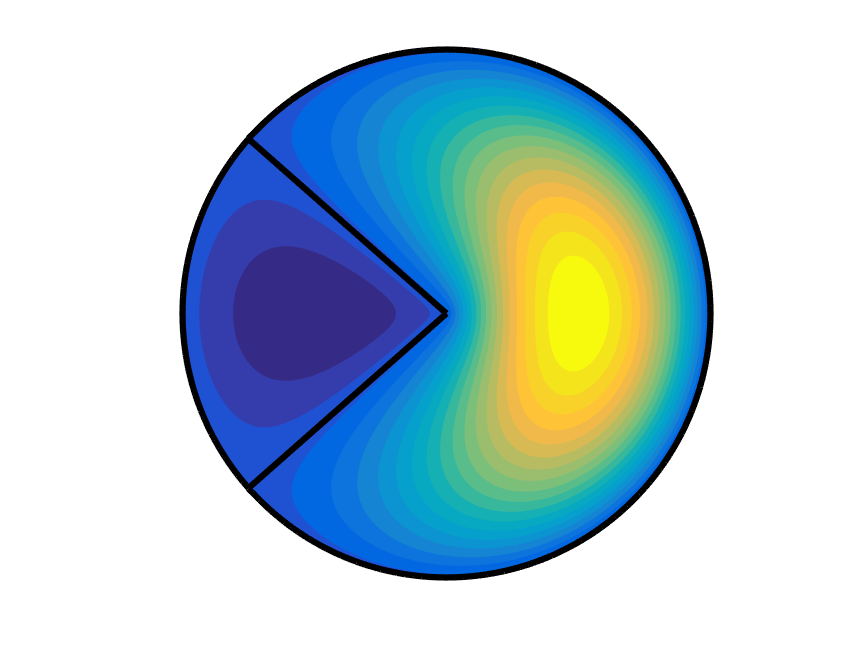}  \put (35,97) {(d)} \end{overpic}
\begin{overpic}[trim={3.0cm 0 2.5cm 0},clip,width=0.13\hsize]{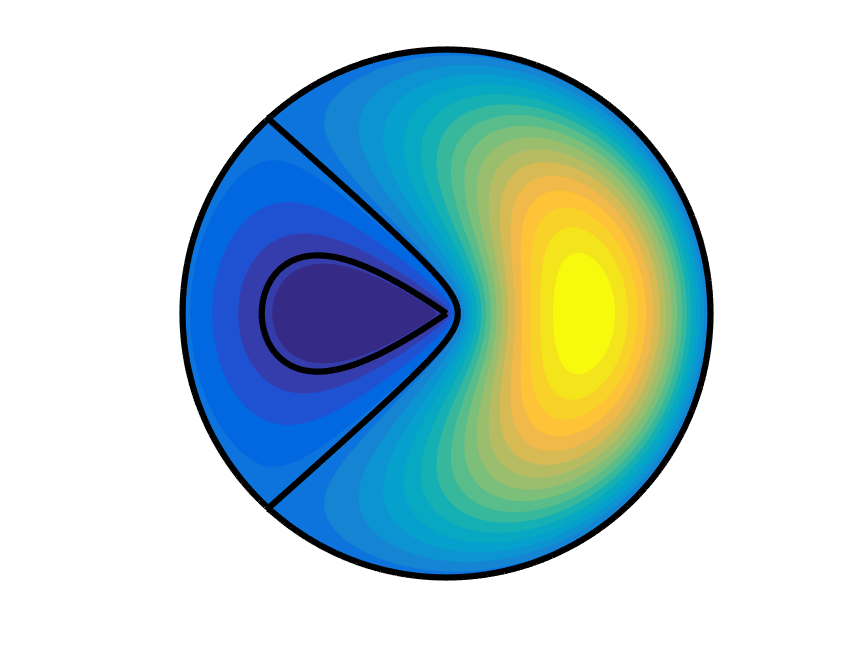}  \put (35,97) {(e)} \end{overpic}
\begin{overpic}[trim={3.0cm 0 2.5cm 0},clip,width=0.13\hsize]{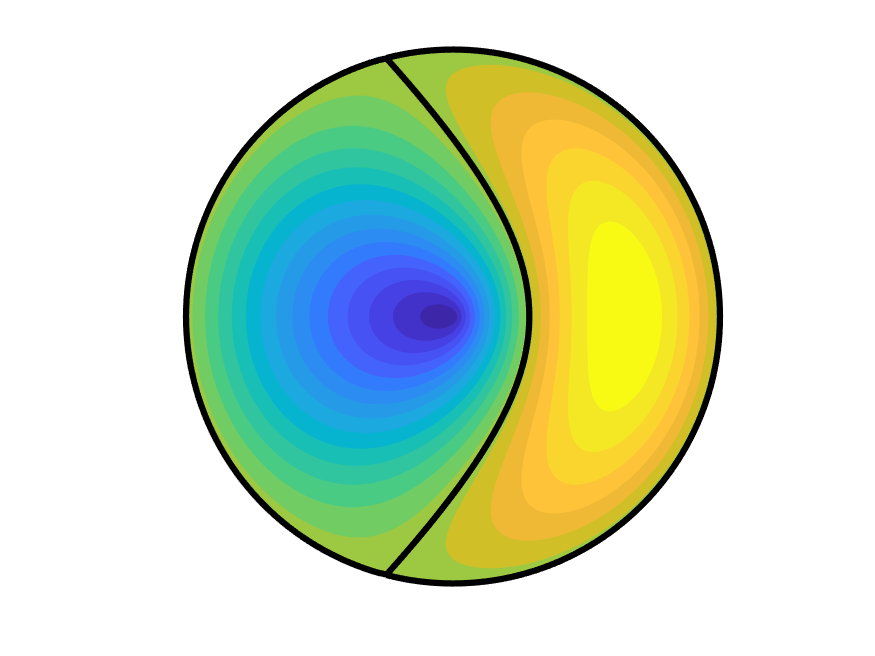}  \put (35,97) {(f)} \end{overpic}
\begin{overpic}[trim={3.0cm 0 2.5cm 0},clip,width=0.13\hsize]{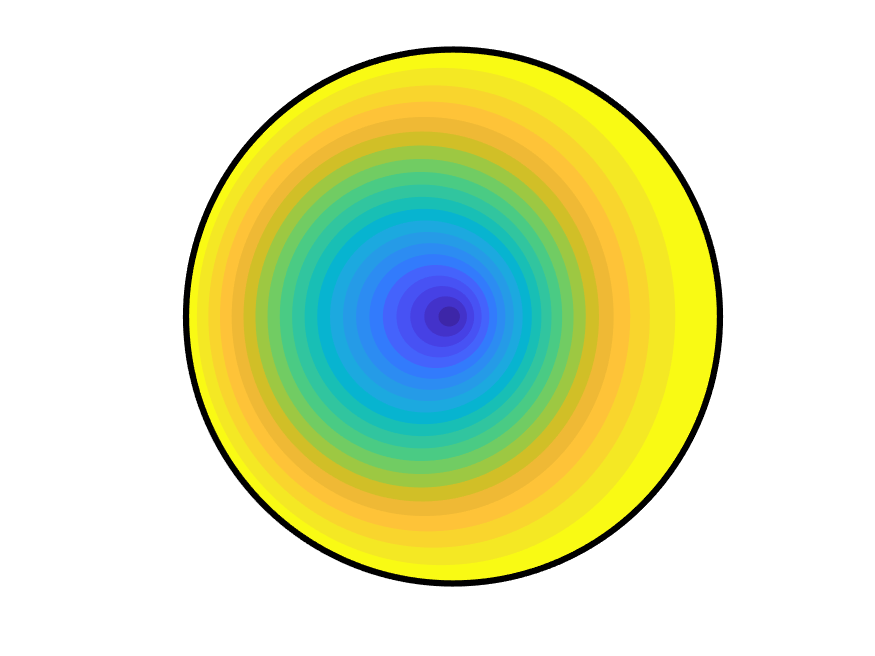}  \put (35,97) {(g)} \end{overpic} \\
\includegraphics[trim={13.5cm 5.5cm 12cm 5cm},clip,width=0.13\hsize]{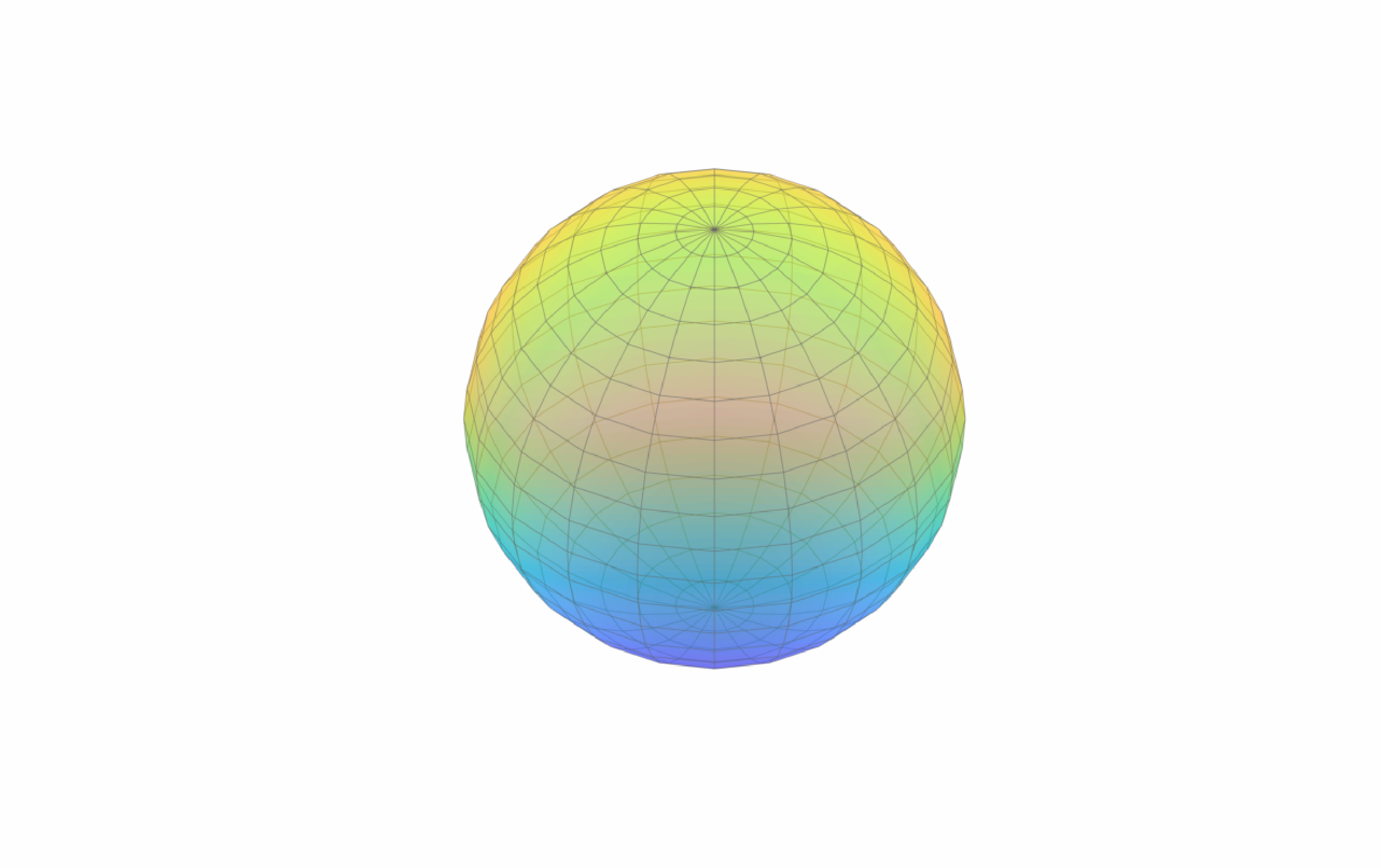} 
\includegraphics[trim={13.5cm 5.5cm 12cm 5cm},clip,width=0.13\hsize]{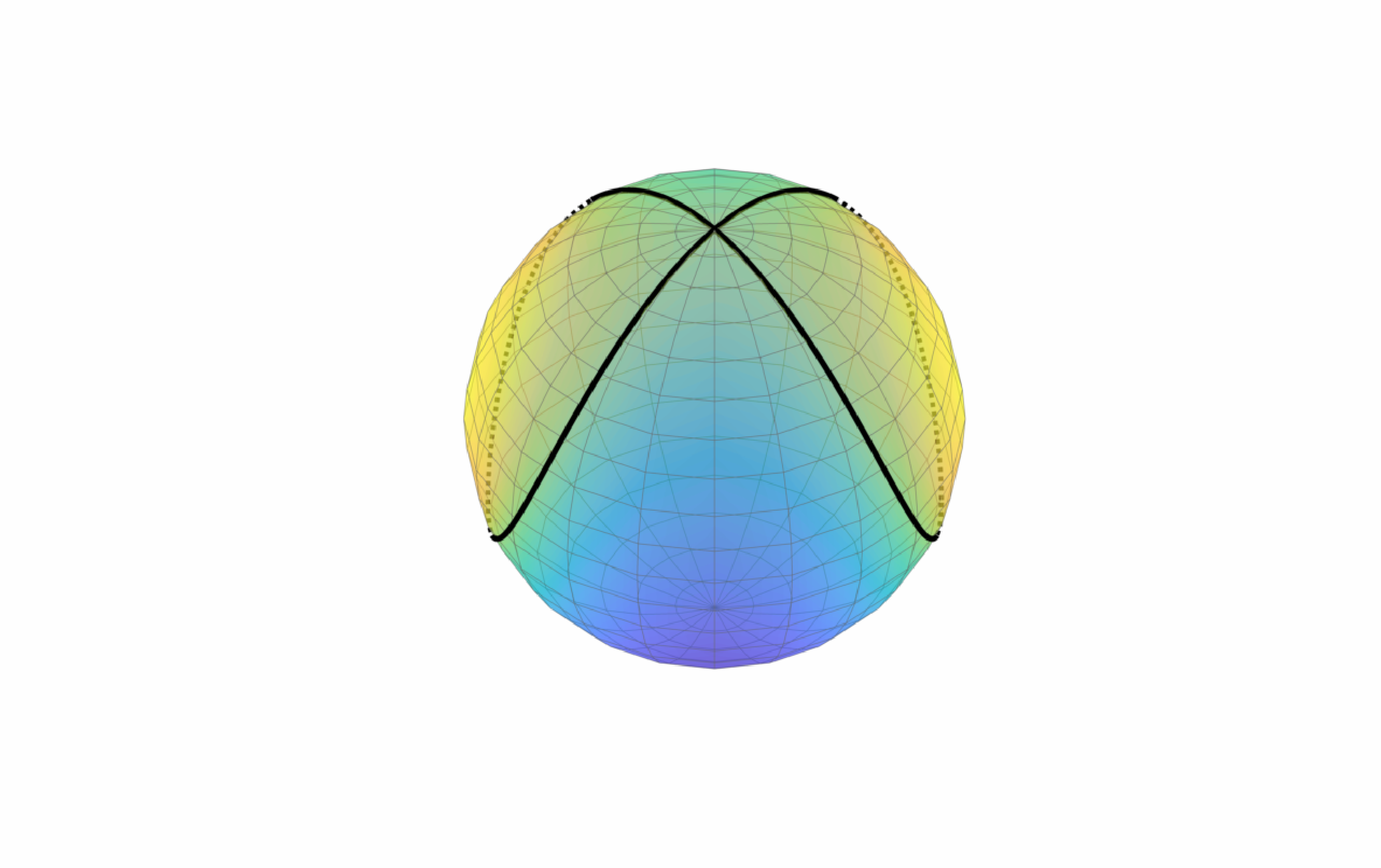} 
\includegraphics[trim={13.5cm 5.5cm 12cm 5cm},clip,width=0.13\hsize]{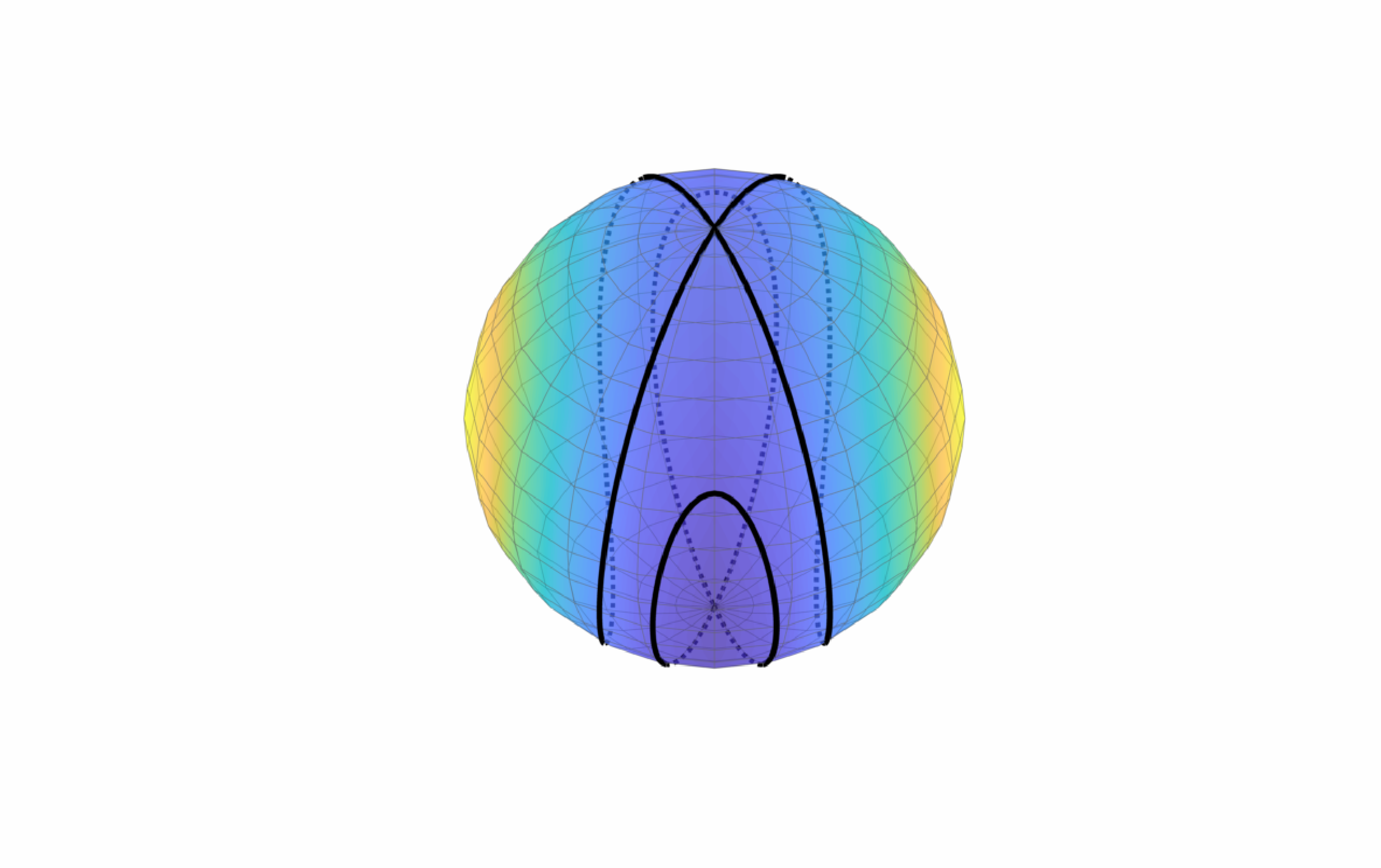} 
\includegraphics[trim={13.5cm 5.5cm 12cm 5cm},clip,width=0.13\hsize]{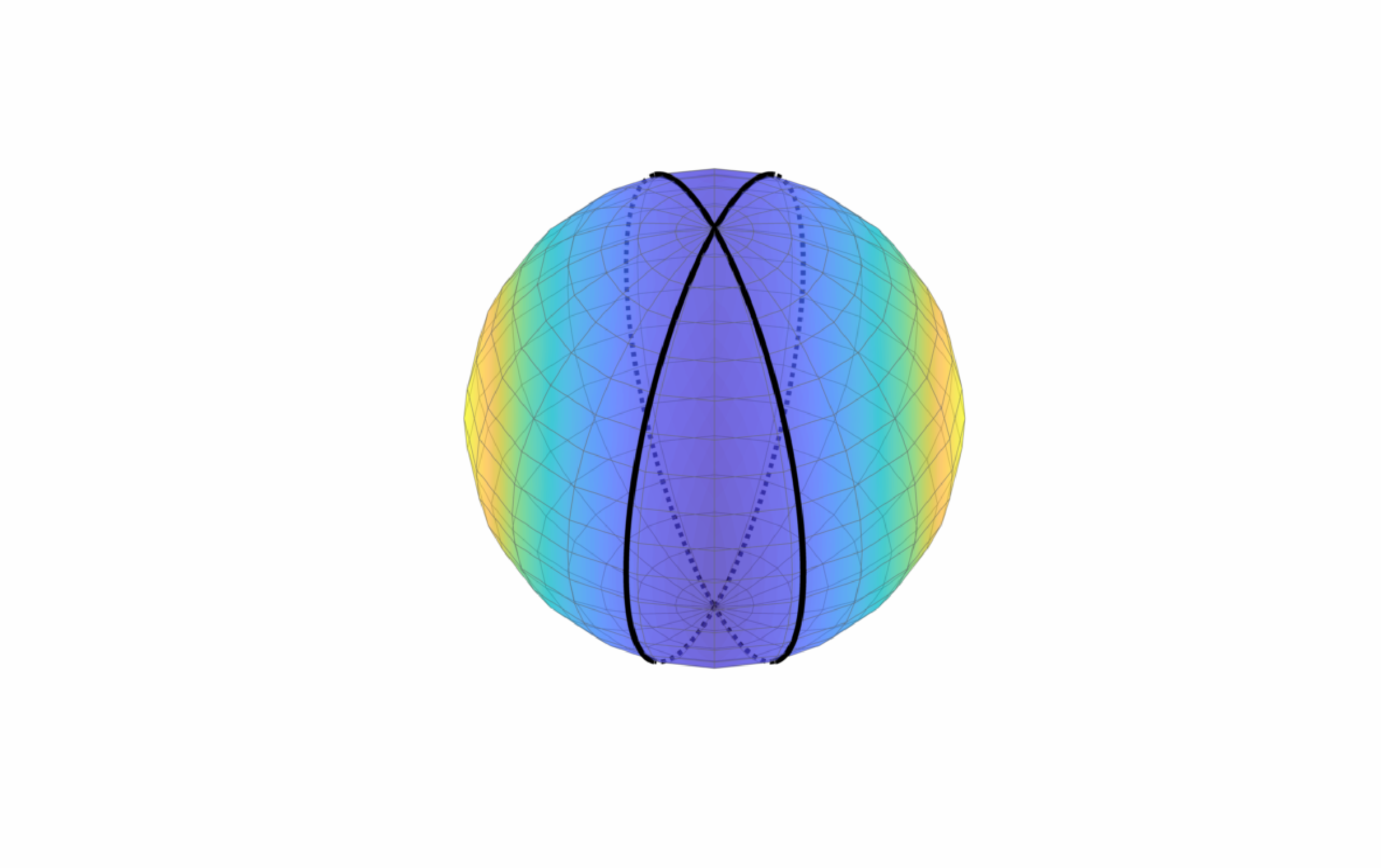} 
\includegraphics[trim={13.5cm 5.5cm 12cm 5cm},clip,width=0.13\hsize]{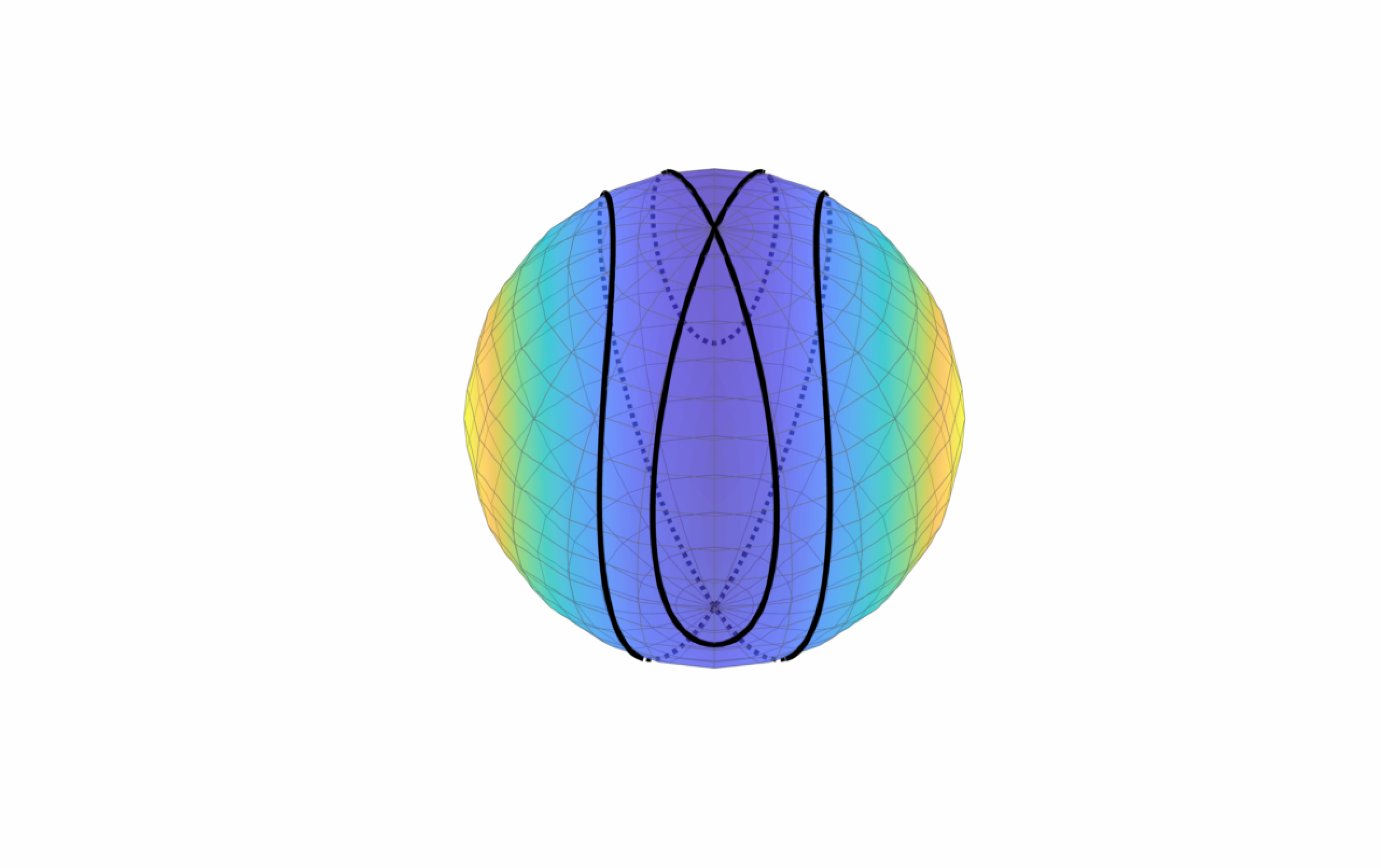} 
\includegraphics[trim={13.5cm 5.5cm 12cm 5cm},clip,width=0.13\hsize]{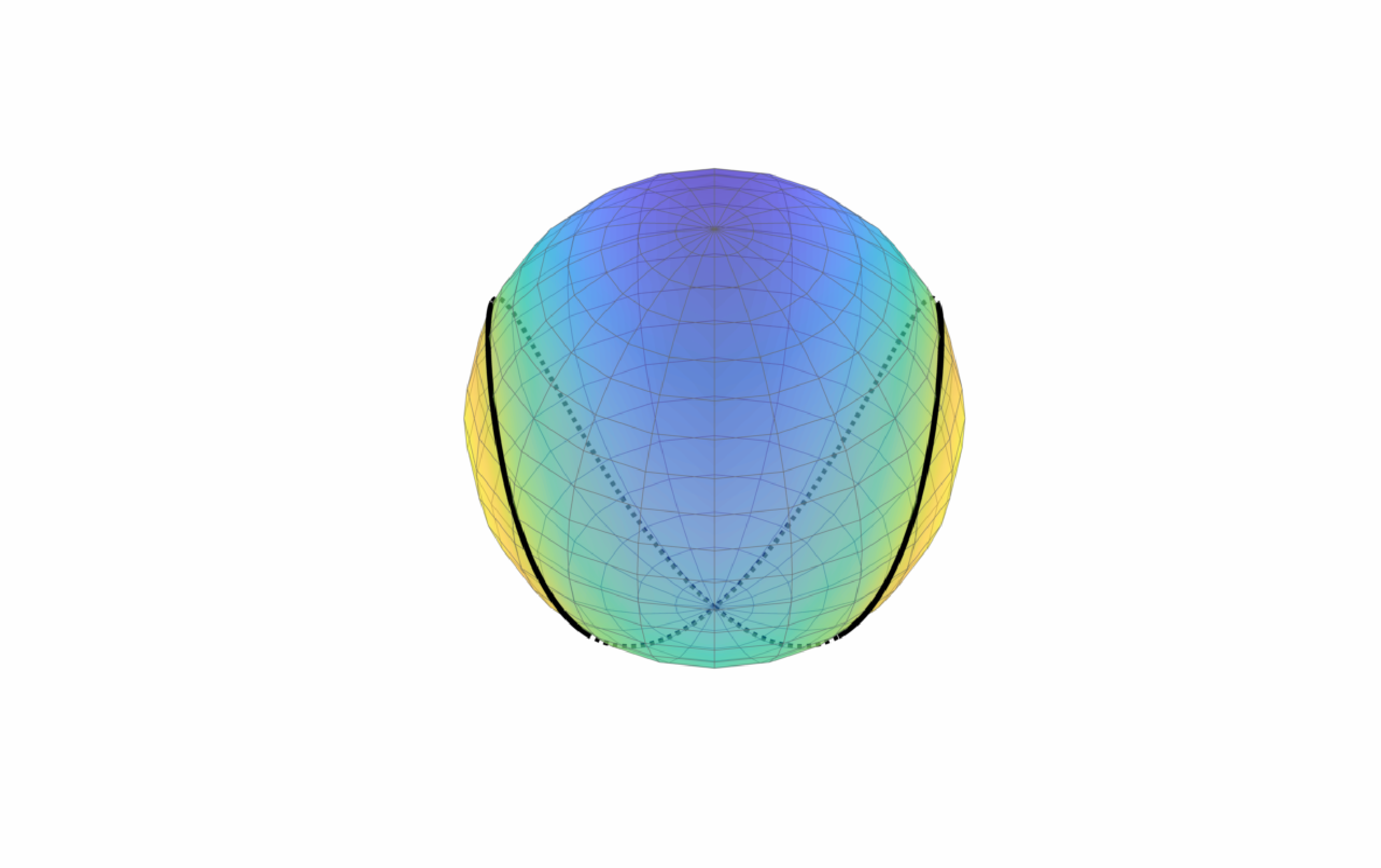} 
\includegraphics[trim={13.5cm 5.5cm 12cm 5cm},clip,width=0.13\hsize]{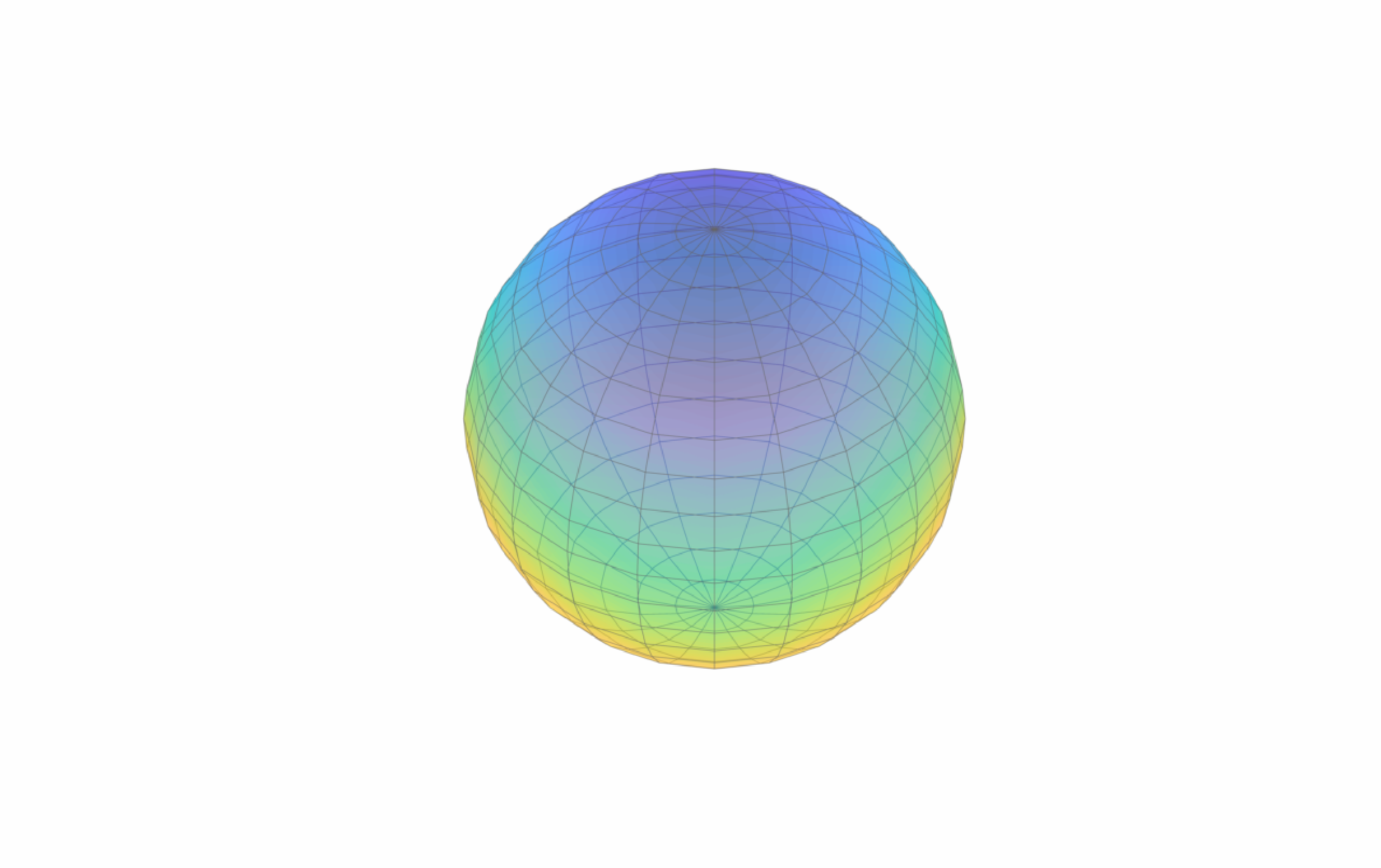} 
\caption{ \label{f2} {\bf The geometry of the projected phase space.} 
Top panels: the $(\varphi,n)$ disk. 
Bottom panels: the ${(\tilde{\varphi},\tilde{n})}$ Bloch-sphere.
The detuning for (a-g) is ${\mathcal{E}/NU=-4/3,-1/3,-0.05,0,0.05,1/3,4/3}$. 
\rmrk{The color stands for the energy of  $\mathcal{H}^{(0)}$, with $M=0$. }
Black lines indicate the separatrices that go through the SPs. 
\hfill 
}
\end{center}
\end{figure*}

\section{Stability, Geometry, and Topology}

\rmrk{Considering the dimer Hamiltonian \Eq{eA0} it is well known that condensation at one orbital is always {\em stable}, 
while condensation at the second orbital becomes {\em unstable} if ${U}$ is large enough. 
This conclusion can be arrived by inspection of \Eq{e2}:   
Without loss of generality let us assume that $U$ is positive (else the energy axis should be flipped); 
Considering condensation in the upper ($-$) orbital, we can regard $\tilde{n} \equiv n_{+}$ 
as the depletion coordinate; Then it follows, using the standard stability analysis of \App{sB}, 
that the North pole (${\tilde{n}=0}$) becomes unstable if ${|\mathcal{E}| < NU}$.}

Considering the trimer Hamiltonian, the supeficial impression is that $\mathcal{H}^{(0)}$ of \Eq{e2} is very similar to \Eq{eA3} of the dimer: 
all we have to do is to rescale the occupation coordinate ${\tilde{n}=2n}$. 
However, the stability analysis of \App{sB} shows that the regime-diagram of \Eq{e2} is in-fact more interesting: 
the condensate (${n=0}$) is unstable for ${ -7NU/6 <  \mathcal{E} <  NU/6 }$,
while the depleted state (${n=N/2}$) is unstable for ${ -NU/6 <  \mathcal{E} <  7NU/6 }$. 
We will focus, in particular, on the the range ${|\mathcal{E}| < (1/6)NU }$, 
where both SPs become unstable. Note that $n=0$ necessary means that $M=0$, while $n=N/2$ is a SP for all $M$ values.

\begin{figure}
\begin{center}

\begin{overpic}[trim={0 2cm 0 -2cm},clip,width=0.3\hsize]{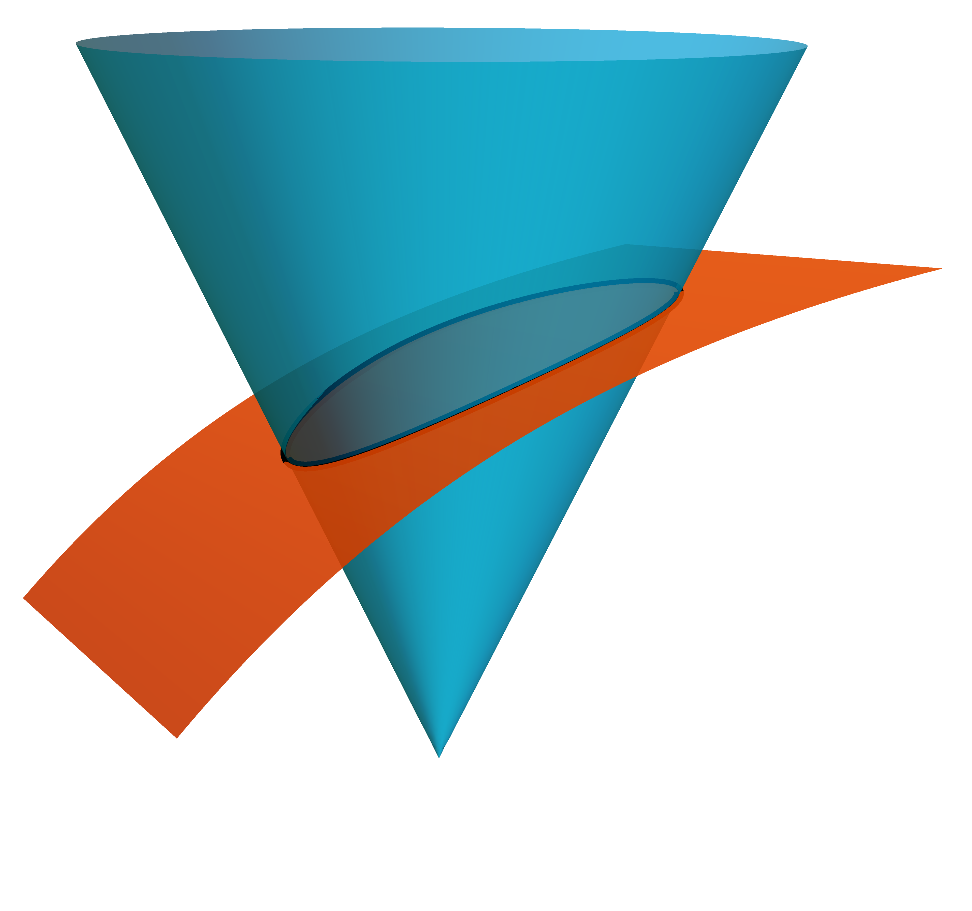} \put (38,88) {(a)} \end{overpic}
\begin{overpic}[trim={0 2cm 0 -2cm},clip,width=0.3\hsize]{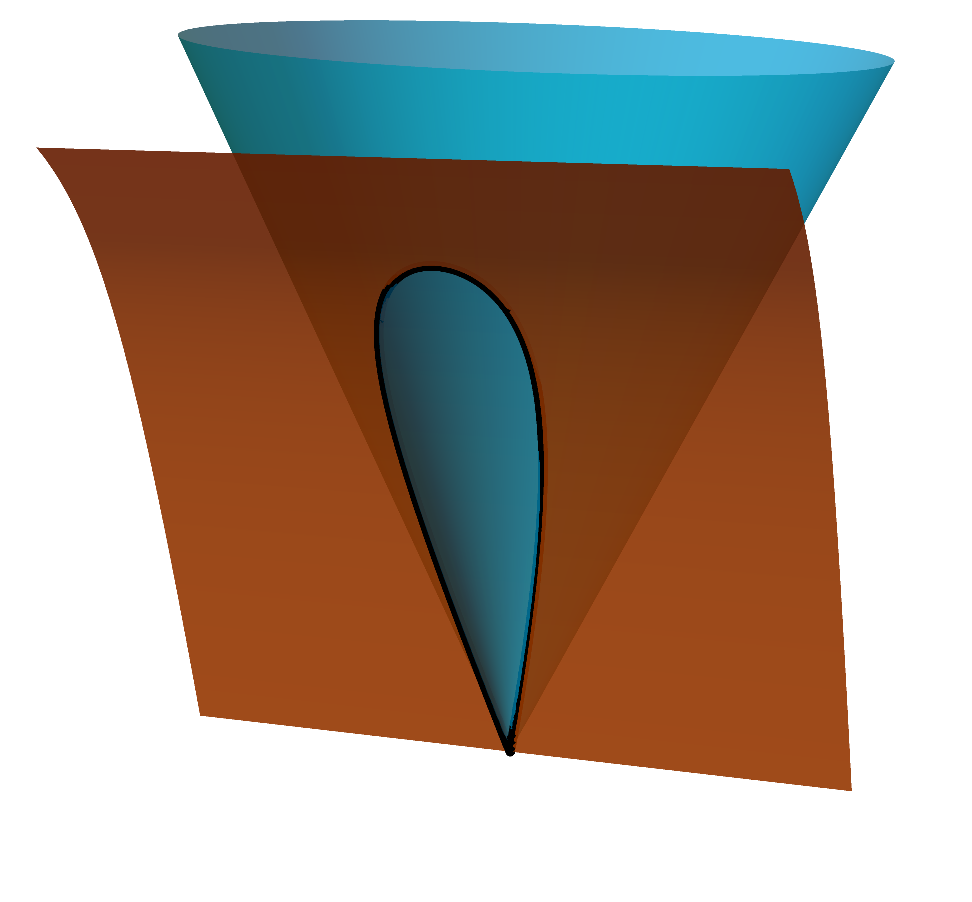} \put (42,88) {(b)} \end{overpic}
\begin{overpic}[trim={5.0cm 0 5cm 0},clip,width=0.3\hsize]{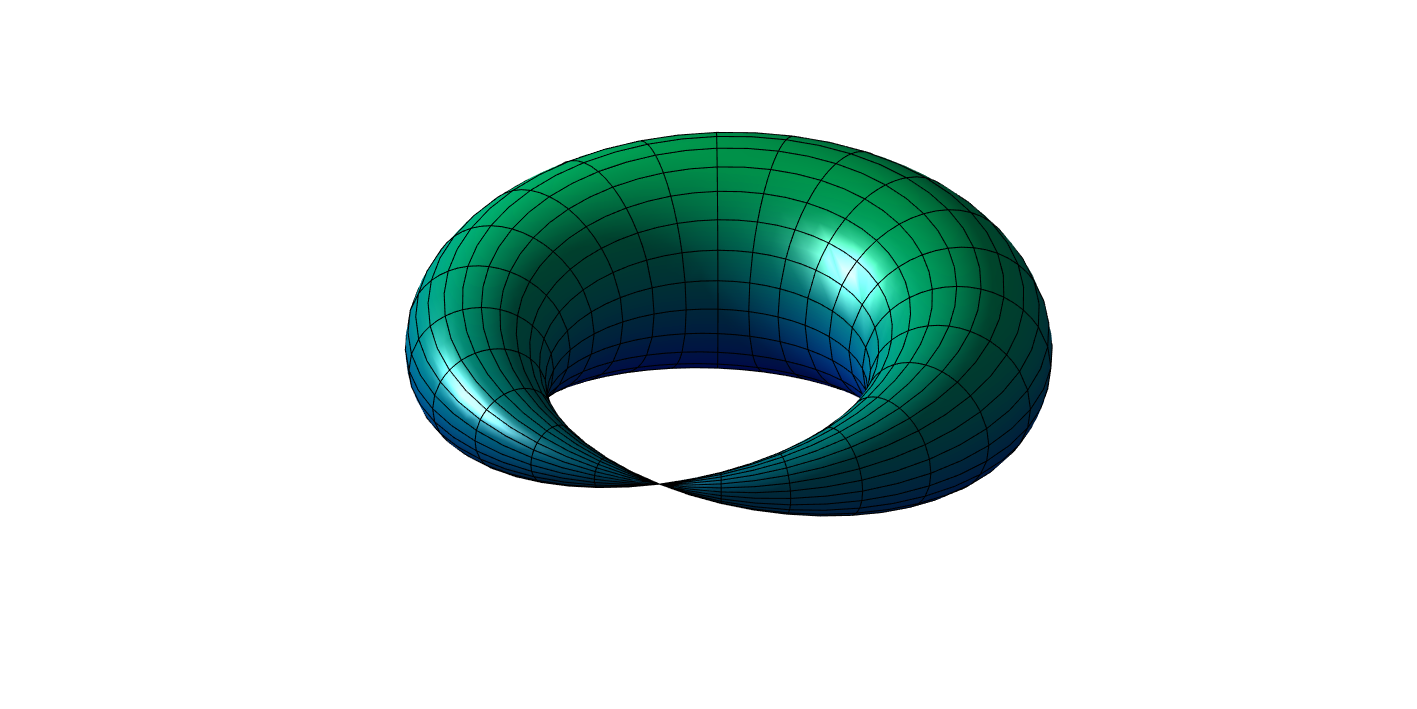}  \put (8,65) {(c)} \end{overpic}
\caption{ \label{f3} 
{\bf Phase space topology}. 
The blue cone is an $M=0$ surface, that intersects with a surface of constant~$E$. 
The existence of an additional coordinate ($\phi$) at each point is implicit.  
The intersection is a torus. Panel~(a) is the typical case, 
while~(b) corresponds to a pinched torus (see text).  
The latter is fully illustrated (with $\phi$) in panel~(c). 
} 
\end{center}
\end{figure}


\sect{Geometry}
The stability analysis reflect the algebraic side of the dynamics, but ignores the geometrical aspect. 
The phase space of the dimer is the Bloch sphere. All the ${(\tilde{\varphi},\tilde{n}{=}0)}$ points are in fact 
the same point, which can be regarded as the North pole of the sphere. 
Same applies to ${(\tilde{\varphi},\tilde{n}{=}N)}$ which can be regarded as the South pole of the sphere. 

But for our 3~site ring \Eq{e2} the geometry of phase-space is different. 
The South pole, it is no longer a single point, 
because different $\varphi$ values indicate different points in phase space. 
So in fact we no longer 
have a Bloch-sphere, but rather we have a Bloch-disc.
Another difference is that 
the angle is folded (${\varphi = 2\tilde{\varphi}}$). 
The phase space structure, for different values of the detuning, is illustrated in \Fig{f2}.
The origin and perimeter of the $(\varphi,n)$ disk should be identified with the North and South poles of the ${(\tilde{\varphi},\tilde{n})}$ Bloch-sphere.
The origin (${n=0}$), if unstable \Fig{f2}(b-e), is the cusp 
on a folded separatrix of half-saddle topography.
The perimeter of the disc is a {\em spread~SP}.
If the spread SP becomes unstable \Fig{f2}(c-f), there is a separatrix 
that comes out from the perimeter in an angle ${ \varphi_{\text{out}} }$, 
and comes back to it in an angle ${ \varphi_{\text{in}} }$. 
Both the approach and the departure from the perimeter along 
the separatrix require an extremely long time.   
We emphasize again that from an algebraic point of view the dynamics 
is the same as if the perimeter were a single point on a Bloch-sphere. 
In the Bloch sphere each phase-space point is duplicated. 
Thanks to this duplication the separatrices that are associated with the SPs
take the familiar figure-8 saddle shape,
which is more illuminating for illustration purpose.

\sect{Topology}
So far we have discussed the one-freedom projected dynamics of ${(\varphi,n)}$. 
But now we have to remember that there is an additional degree of freedom ${(\phi,M)}$. 
We consider the dynamics that is generated by $\mathcal{H}^{(0)}$, where~$M$ 
is a constant of motion, and the conjugate angle is doing circles 
with ${\dot{\phi} =  \partial \mathcal{H}^{(0)} / \partial M}$.
A trajectory that is generated by $\mathcal{H}^{(0)}$ covers a torus in phase space. 
A useful way for visualizing the tori is based on the $SU(1,1)$ symmetry \cite{SU11,PhysRevA.83.033605} of $\mathcal{H}^{(0)}$. The ${(\varphi,n)}$ dynamics is the intersection of constant~$E$ and constant~$M$ surfaces, see \Fig{f3} and \App{sC}. In particular the ${M=0}$ surface is a cone, whose tip corresponds to ${n=0}$, while its outer boundary to ${n=N/2}$.
If the intersection forms a closed loop, as in \Fig{f3}a, the trajectory covers a torus in phase space. But if the trajectory goes through ${n=0}$, as in \Fig{f3}b, we get a {\em pinched torus}, see \Fig{f3}c. This is because the $\phi$-circle at ${n=0}$ has zero radius. 
This ``zero radius" is explained as follows: if ${n=0}$ then necessarily ${n_1=n_2=0}$, 
hence all the ${(\varphi,\phi)}$ angles degenerate, representing a single phase-space point.  
\rmrk{In the projected dynamics \Fig{f2}, the cusped trajectory which goes through $n=0$ (when unstable) is merely a projection of the pinched torus.}

\sect{Definition of $\beta$} 
\rmrk{Consider a trajectory that has a period $T$ in ${(\varphi,n)}$. For illustration this can be the trajectory that loops along the intersection in \Fig{f3}a.  Clearly, this trajectory is in general not a closed loop in the full phase space representation. Rather it winds on a two-dimensional torus. We define $\beta $ as the change in $\phi$ during time~$T$. For a trajectory that passed through an unstable SP we have ${T \rightarrow \infty}$, and~${\beta}$ is ill-defined. In \Fig{f1}a, we plot $\beta$ as a function of $M$ and $E$.}

\section{The swap transition}

Recall that $\mathcal{E}$ is controlled experimentally by the rotation frequency of the device. \Fig{f2} shows the projected dynamics for different values of the detuning $\mathcal{E}$. \rmrk{In panels (c-e) both SPs are unstable, and we see how they {\em swap} as the detuning changes sign. At the transition the two separatrices coalesce, thus forming connection between the origin $n=0$ (which supports the condensate) and the perimeter $n=N/2$ (where the ${k{=}0}$ orbital is completely depleted).}

On the Bloch-sphere, both North and South poles, when unstable, take the familiar 8-like shape. 
As we previously argued, this is due to the fact that the phase-space is duplicated, and that all the $\varphi$ values at the South pole are regarded as a single point. This physically unfaithful presentation possibly better reflects what do we mean by ``swap of separatrices".  
We note that the Poincare sections in \cite{sfc}, that had been presented before we gained proper understanding
of the swap-transition, were physically unfaithful is the same sense.

\begin{figure}
\begin{center}
\begin{overpic}[trim={4.0cm 1cm 3cm 0.5cm},clip,width=0.3\hsize]{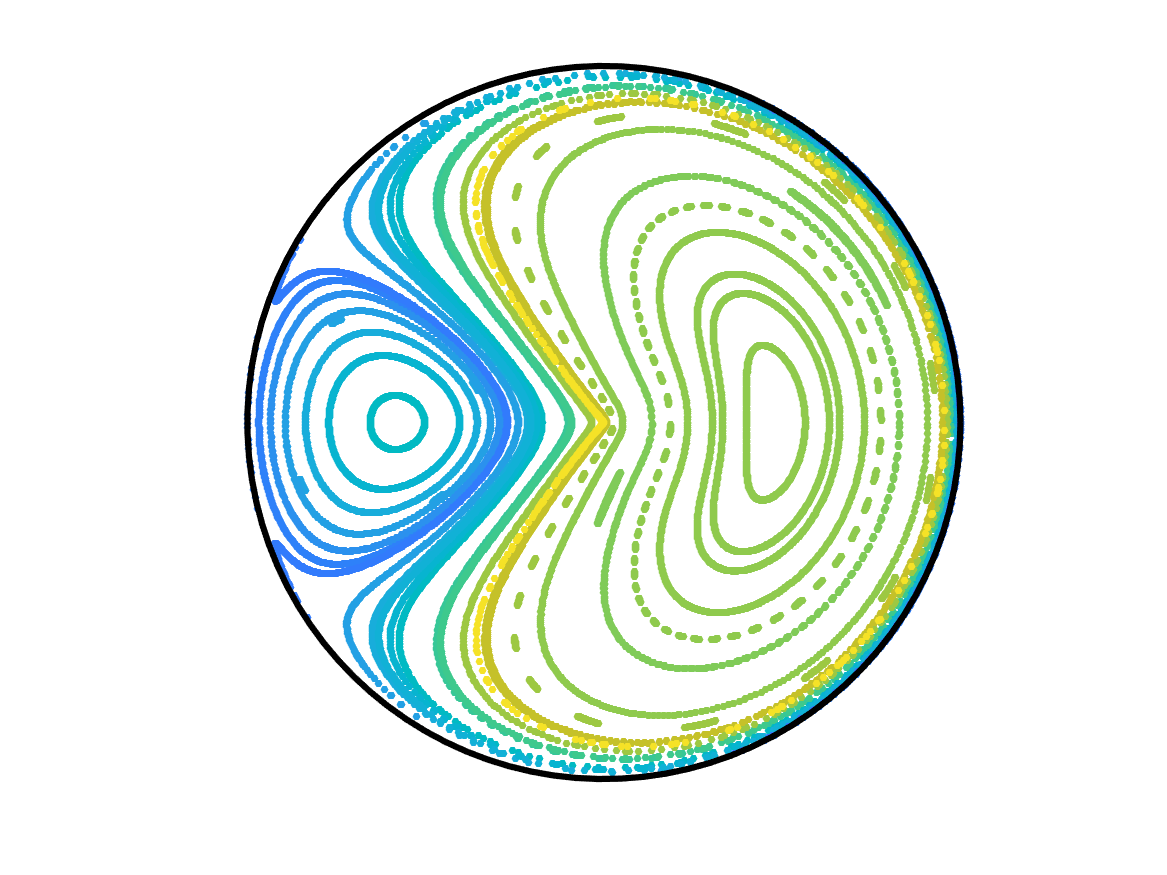}  \put (4,92) {(a)} \end{overpic}
\begin{overpic}[trim={4.0cm 1cm 3cm 0.5cm},clip,width=0.3\hsize]{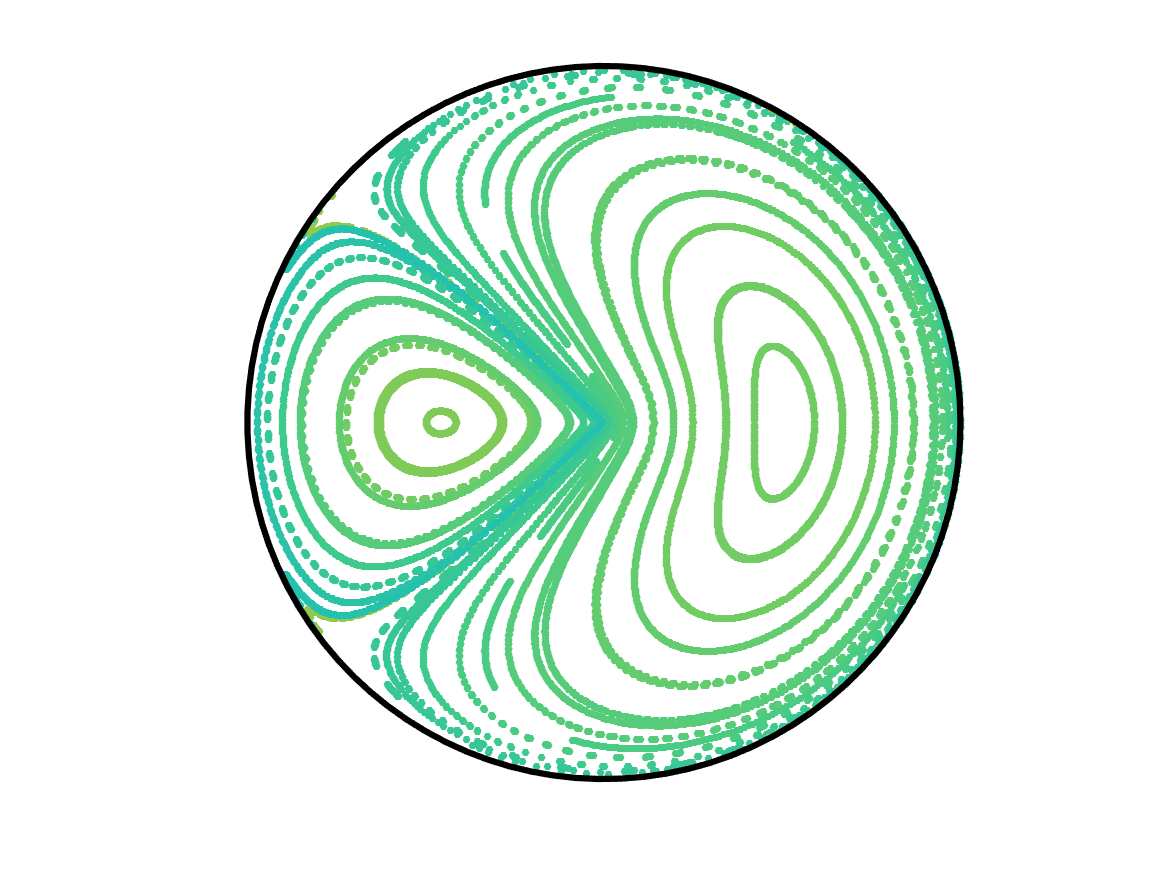}  \put (4,92) {(b)} \end{overpic}
\begin{overpic}[trim={4.0cm 1cm 3cm 0.5cm},clip,width=0.3\hsize]{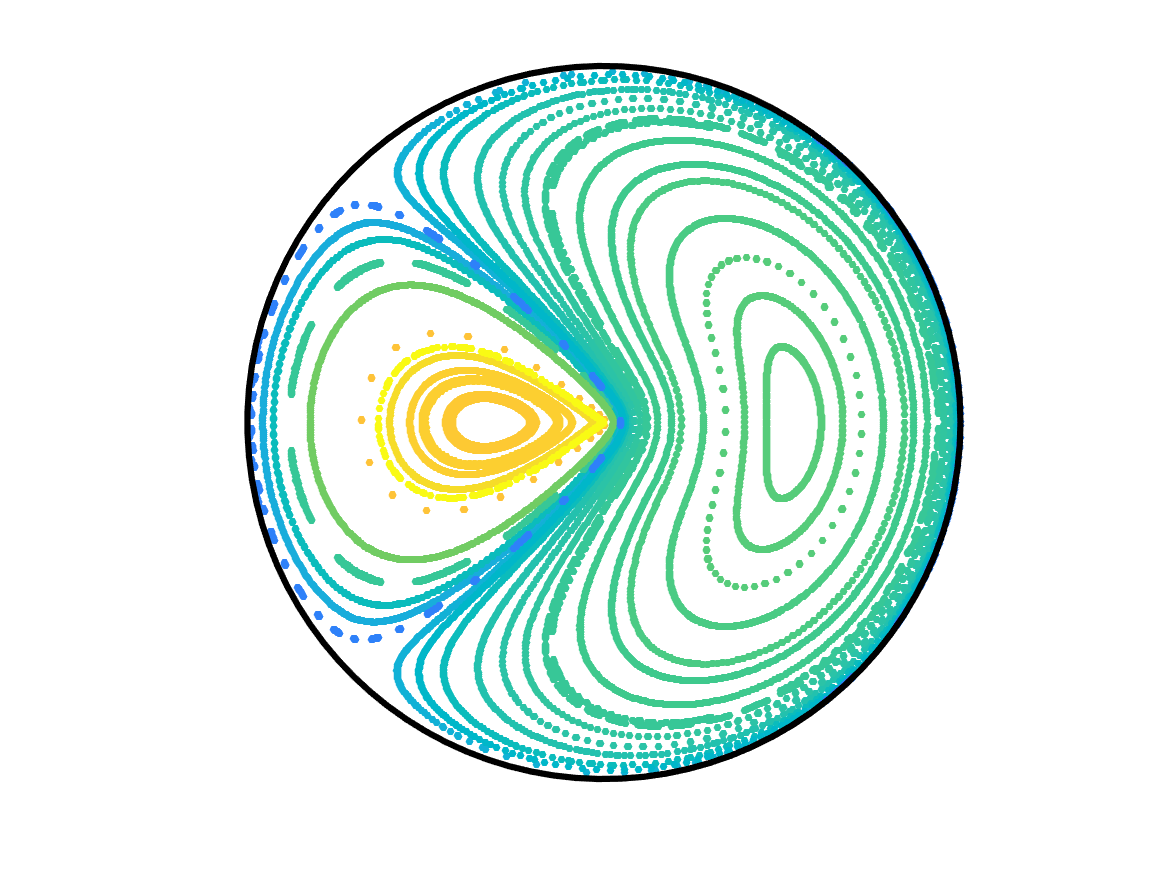}  \put (4,92) {(c)} \end{overpic}
\begin{overpic}[trim={4.0cm 1cm 3cm 0.5cm},clip,width=0.3\hsize]{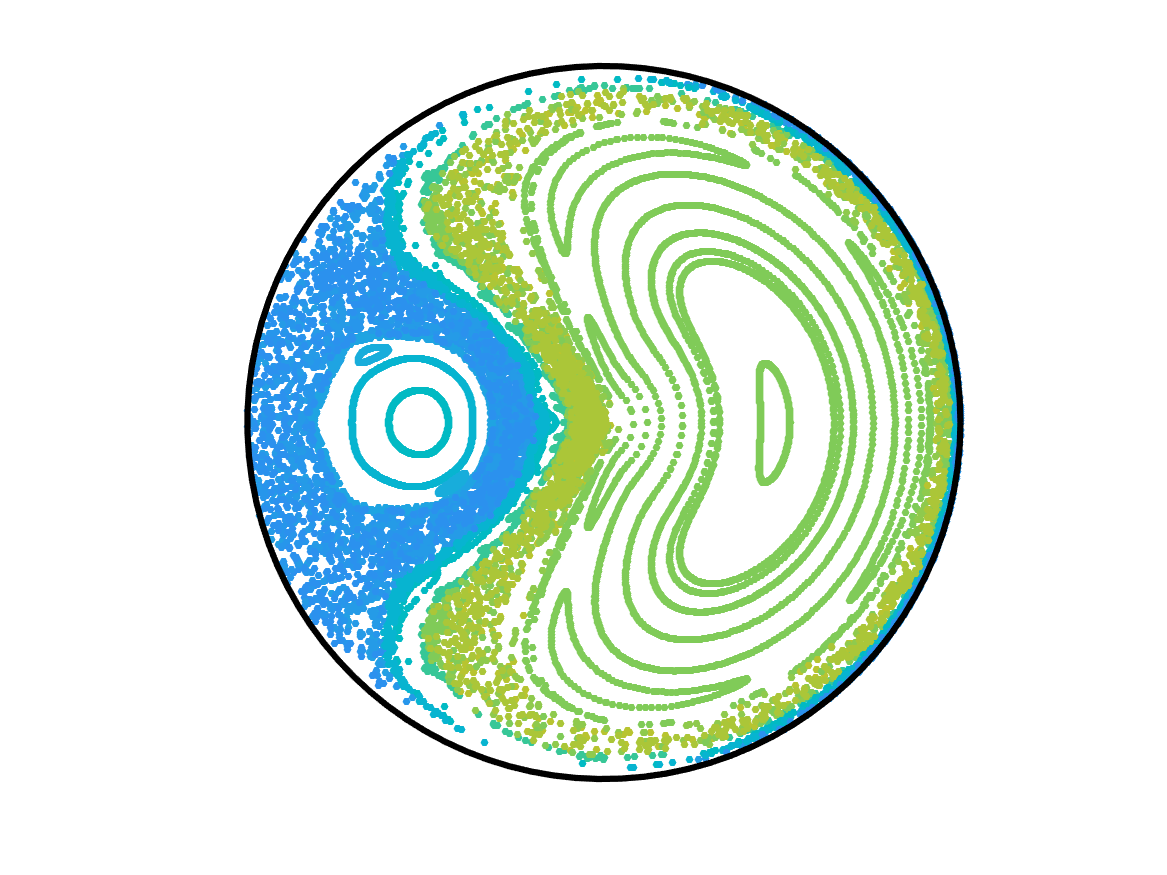}  \put (4,92) {(d)} \end{overpic}
\begin{overpic}[trim={4.0cm 1cm 3cm 0.5cm},clip,width=0.3\hsize]{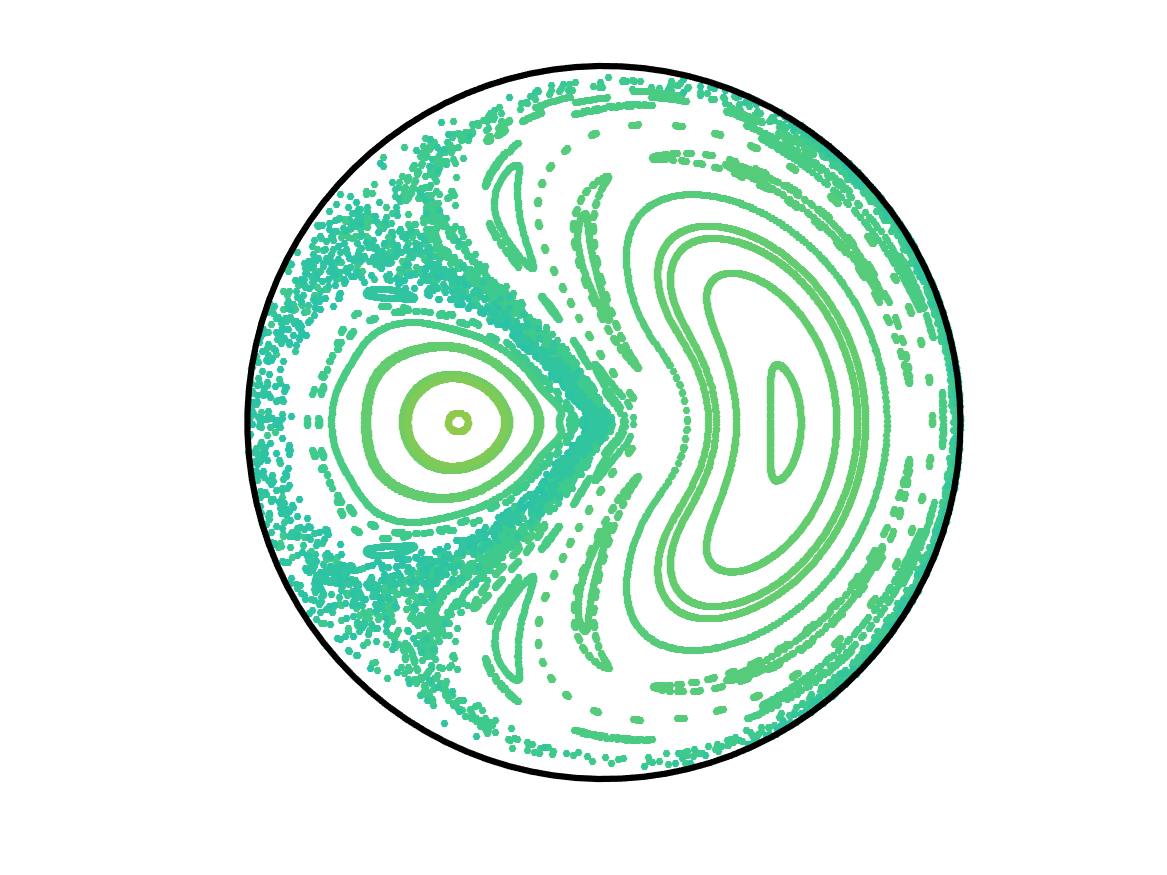}  \put (4,92) {(e)} \end{overpic}
\begin{overpic}[trim={4.0cm 1cm 3cm 0.5cm},clip,width=0.3\hsize]{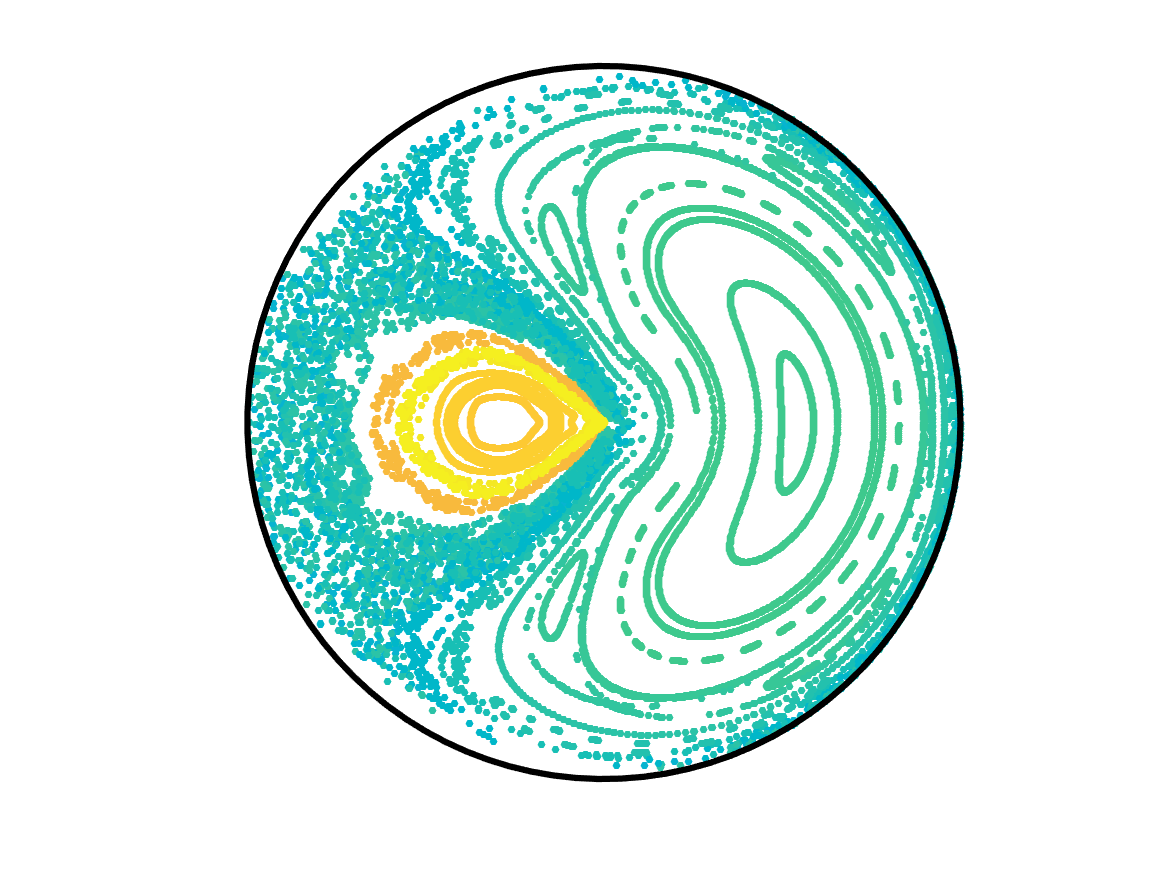}  \put (4,92) {(f)} \end{overpic}

\caption{ \label{f4} 
{\bf Poincare sections.} 
The dynamics of the full Hamiltonian \Eq{e1} projected to the $(\varphi,n)$ disk.
All the trajectories are launched with the same energy as that of the condensate, 
and the section is chosen to be $q_2=0$.  
The left to right arrangement of the panels is by detuning ${\mathcal{E}/NU=-0.05,0,0.05}$, 
in one-to-one correspondence with \Fig{f2}(c-e).
In the upper panels the interaction strength is ${NU \sim 1}$, in units of the BHH hopping frequency K, 
while in the lower panels it is doubled, keeping $\mathcal{E}/NU$ fixed. 
The color-code (from yellow to blue) corresponds to the trajectory-averaged occupation $n$ (from $N/8$ to $N/2$). 
\hfill
} 
\vspace*{5mm}
\begin{overpic}[width=0.32\hsize]{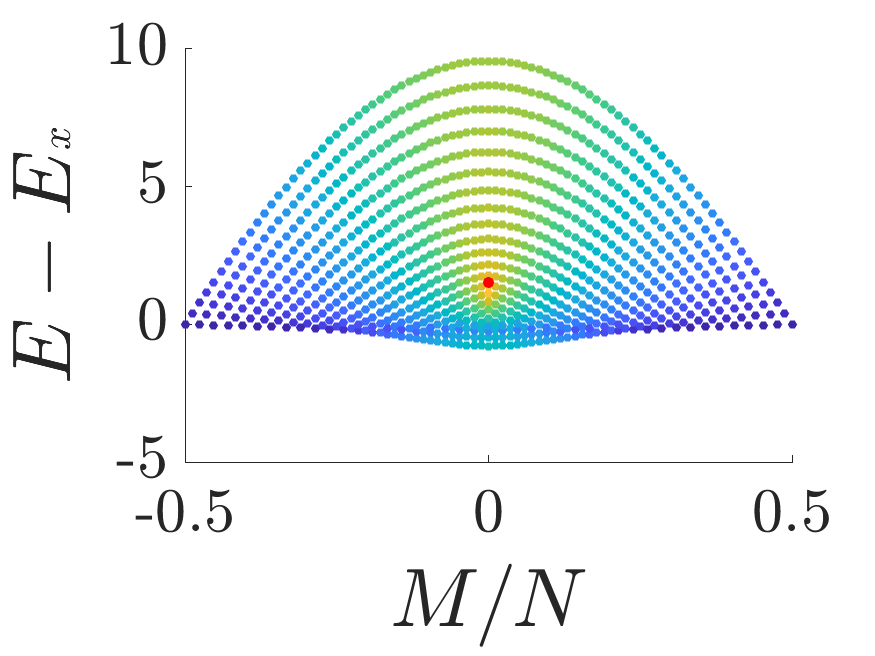}  \put (20,62) {(a)} \end{overpic}
\begin{overpic}[width=0.32\hsize]{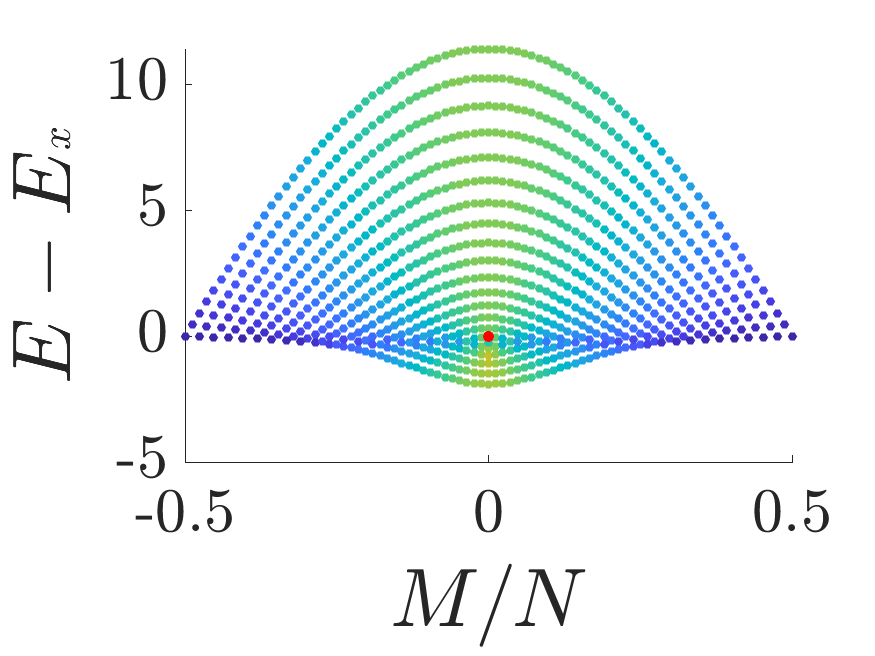}   \put (20,62) {(b)} \end{overpic}
\begin{overpic}[width=0.32\hsize]{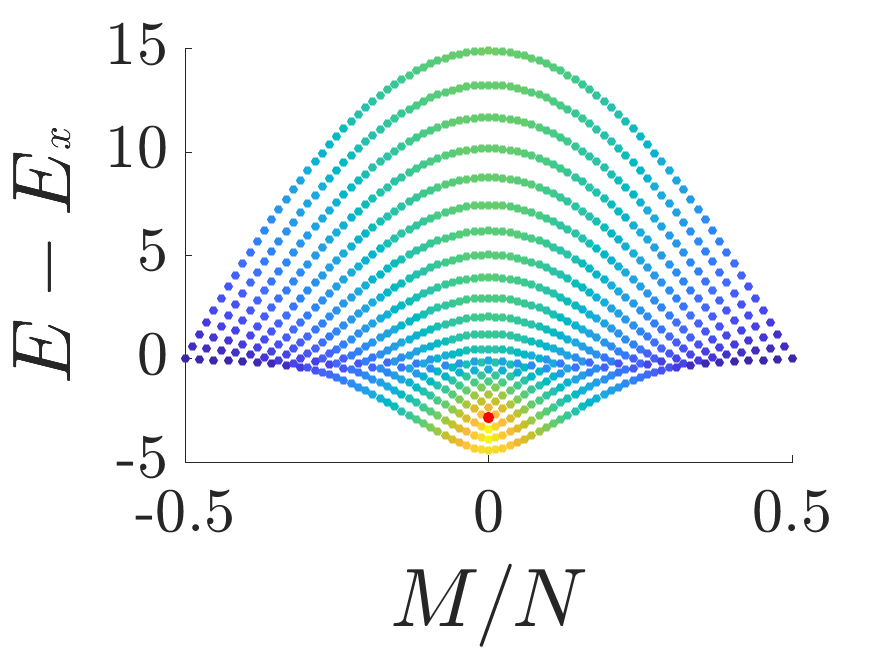}  \put (20,62) {(c)} \end{overpic}
\begin{overpic}[width=0.32\hsize]{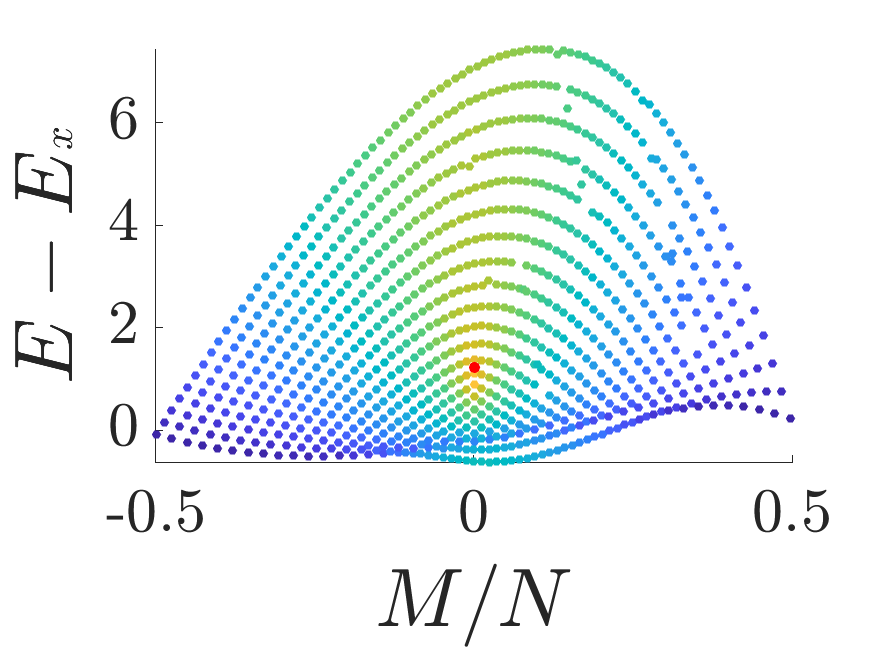}   \put (20,62) {(d)} \end{overpic}
\begin{overpic}[width=0.32\hsize]{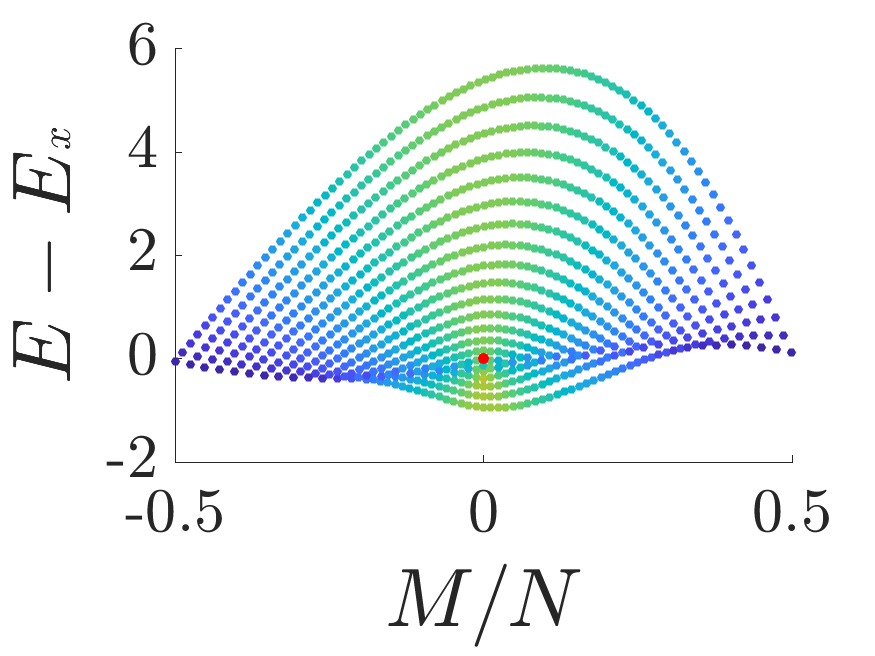}  \put (20,62) {(e)} \end{overpic}
\begin{overpic}[width=0.32\hsize]{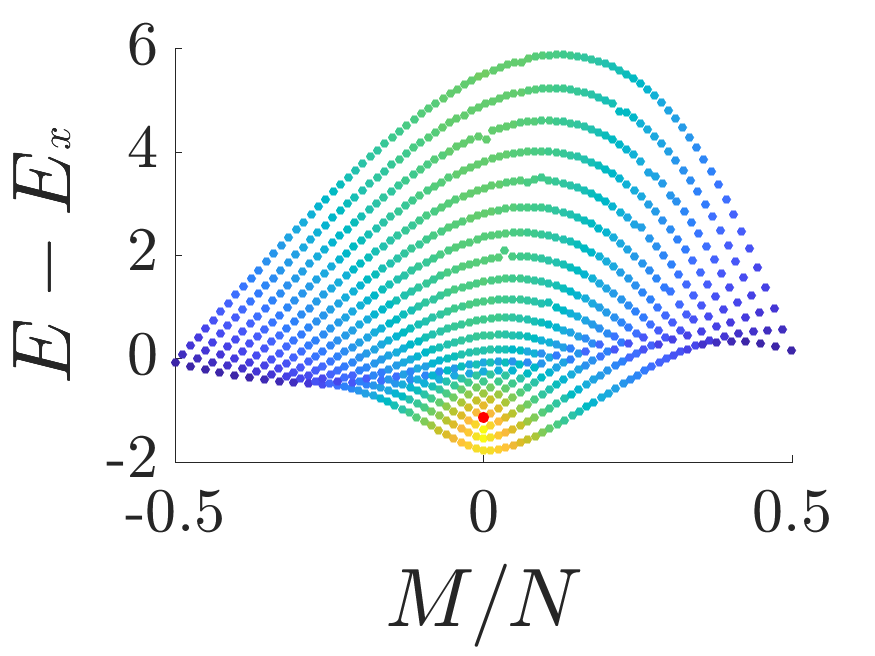}  \put (20,62) {(f)} \end{overpic}
\begin{overpic}[width=0.32\hsize]{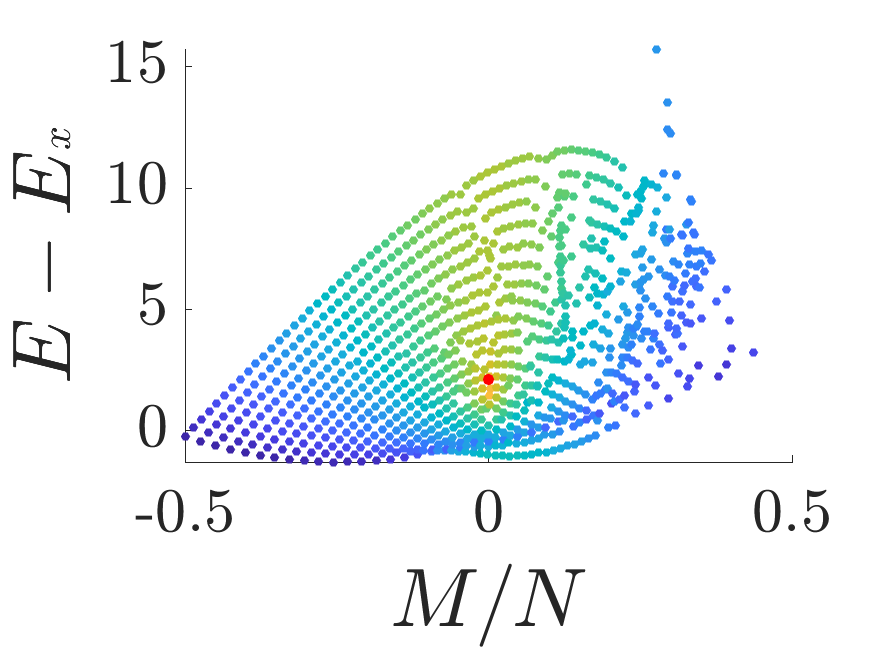}  \put (20,62) {(g)} \end{overpic}
\begin{overpic}[width=0.32\hsize]{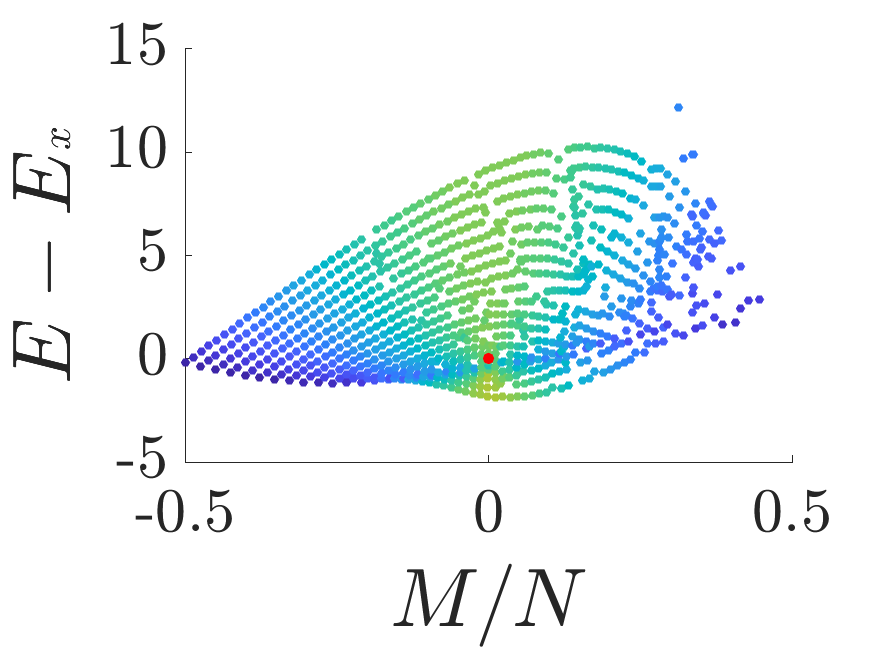}  \put (20,62) {(h)} \end{overpic}
\begin{overpic}[width=0.32\hsize]{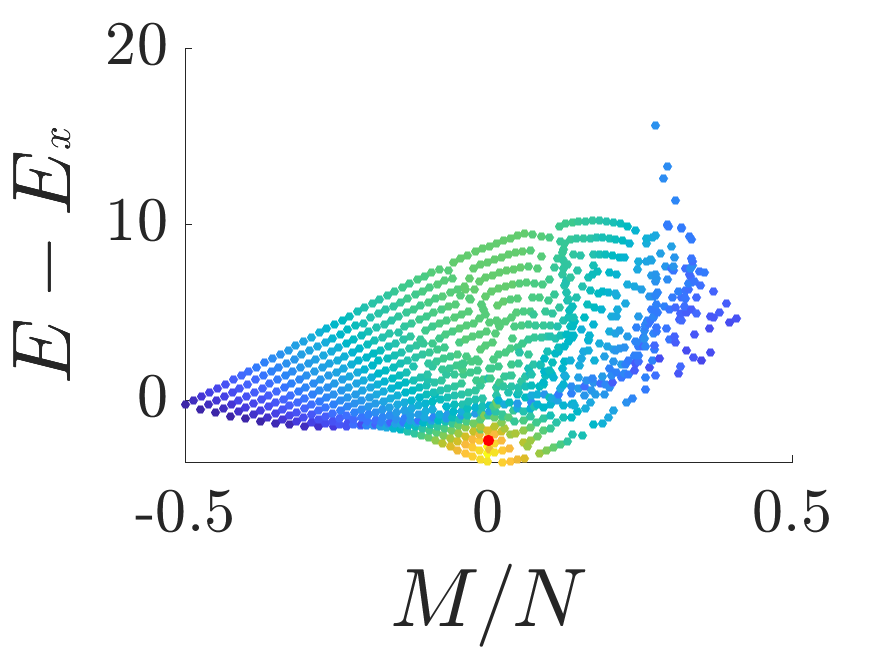}  \put (20,62) {(i)} \end{overpic}
\caption{ \label{f5} 
{\bf The spectrum.} 
The left to right arrangement of the panels is as in \Fig{f4}.  
In the upper row we plot the spectrum of $\mathcal{H}^{(0)}$, 
while the two other rows provided the spectrum of $\mathcal{H}$
in one-to-one correspondence with \Fig{f4}.   
All the spectra refer to ring with $N{=}42$ particles. 
The points are colored by the expectation value of $n$, 
with the same colorcode as in \Fig{f1}b and \Fig{f4}.
\hfill
} 
\end{center}
\end{figure}

Once the $\mathcal{H}^{(\pm)}$ terms are added, a connecting quasi-stochastic strip is formed, through which the initial state can decay. This is shown in \Fig{f4}, where we plot a Poincare section of the full Hamiltonian \Eq{e1}.
One should note the subtle relation between the perspective of \Fig{f4} and that of \Fig{f2}. A panel of the  latter displays sections of ${M=0}$ tori that form a vertical subset in a \Fig{f1}-type ${(M,E)}$ diagram, while a panel of \Fig{f4} displays sections of same~$E$ trajectories that form a horizontal subset of such diagram. The pinched torus is contained in both subsets. 

Away from the swap transition, the chaotic region around ${n=0}$ is bounded by the surviving Kolmogorov-Arnold-Moser (KAM) tori, forming a {\em chaotic pond} which is isolate from the perimeter region. Hence the depletion of the condensate is arrested. It is only in the vicinity of the swap transition that a connected chaotic pathway to depletion is formed. Thus, a local stability analysis of the SP using the standard Bogoliubov procedure does not provide the proper criterion for superflow metastability.

\section{Quantization}
The classical structure of phase-space is reflected in the many-body spectrum. 
If chaos is ignored the eigenstates can be labeled by the good quantum numbers 
that are determined by the commuting operators ${M}$ and ${\mathcal{H}^{(0)}}$, as in \Fig{f1}b.
If we add the $\mathcal{H}^{(\pm)}$ terms we can still order the energies 
according to the expectation value ${\braket{M}}$. 
Several examples are provided in \Fig{f5}.
For presentation purpose, the perimeter energy ${E_x(M)}$, 
which corresponds to maximum depleted state (${n=N/2}$),  
is taken as the reference. 
%
%
%
\Fig{fs3} of \App{sD} provides spectra 
for the whole range of the detuning parameter, 
corresponding to the phase space plots of \Fig{f2}.

From a semiclassical perspective, if we ignore the chaos, 
each point can be associated with an EBK torus \App{sE}.
Namely, the ``good quantum numbers" are quantized values of the action variables. 
The lattice arrangement of the energies in \Fig{f1}b  
reflects the way that the tori are embedded in phase-space, 
while the chaos, once added, blurs it locally, see \Fig{f5}.  
This lattice arrangement is supported by a classical skeleton 
that is formed  by a pinched torus (marked by a red dot), 
and an ${E=E_x(M)}$ separatrix. 
At the vicinity of the separatrix the spectrum is dense, 
reflecting that the frequency of the motion goes to zero.
Irrespective of that, the quantum spectrum has a topological defect 
that is described by a monodromy (to be further discussed below). 
This monodromy reflects the presence of the pinched torus.      
The sequence of panels in \Fig{f5} shows how the swap transition 
is reflected in the quantum spectrum.
This transition happens as the red dot, 
which corresponds to the pinched torus, 
crosses the $E=E_x$ separatrix line. 
We see how the yellow condensation region is diminished at the transition.

\section{Monodromy calculation}
The concept of monodromy is pedagogically summarized in \App{sE}. 
For our model system, in the absence of chaos, we have in involution 
the generators ${H_1 = \mathcal{H}^{(0)}}$ and ${H_2 = M}$. 
The trajectories that are generated for a given $E$ and $M$ form a torus.
Any point on the torus is accessible by generating a walk of duration ${(t_1,t_2)}$.
Consider the projected dynamics in ${(\varphi,n)}$. 
A given trajectory has a period $T$, 
but in the full phase-space it is, in general, not periodic, 
because $\phi$ has advanced some distance $\beta$. 
It follows that in order to get a periodic walk on the torus,
the ${t_1=T}$ evolution that is generated by ${H_1}$, 
should be followed by a ${t_2=-\beta}$ evolution that is generated by ${H_2}$. 
The so called rotation angle, $\beta$, characterizes the torus,
and is imaged in \Fig{f1}a. 
Note that a ${t_2=4\pi}$ evolution that is generated by ${H_2=M}$ 
is a periodic trajectory in phase space, 
because it does not affect the ${(\varphi,n)}$ degree of freedom. 
\rmrk{We conclude that the set of periodic walks 
forms a lattice that can be spanned by the basis vectors}
\be{4}
\vec{\tau}_{a} &=& (T,-\beta) \\ 
\vec{\tau}_{b} &=& (0,4\pi) 
\label{e5}
\eeq


A remark is in order regarding the determination of the $4\pi$ in \Eq{e5}.
It should be clear that the original phases ${(\varphi_0, \varphi_1,\varphi_2)}$ 
are defined ${ \mod(2\pi) }$. 
Next we define the coordinates ${q_1=\varphi_1-\varphi_0}$ and ${q_2=\varphi_2-\varphi_0}$, 
and the alternate coordinates ${\phi=q_1-q_2}$ and ${\varphi=q_1+q_2}$.
If the alternate coordinates are regarded as ${ \mod(2\pi) }$ angles, 
it follows that each ${(\varphi, \phi)}$ represent {\em two} points in $q$ space, 
and each ${(\varphi,n)}$ in our sections is the projection of a $4\pi$ circle.
Consider a trajectory that is generated using ${H_2=M}$. 
In the ${(\varphi_1,\varphi_2)}$ torus it will have a constant ${\varphi}$. 
You will have to run ${t}$ a ${4\pi}$ interval in order to get back to the starting point.

\sect{Quantum to classical duality}
Let us now go back to \Fig{f1}a, where we plot $\beta$ as a function of $M$ and $E$. 
One can immediately spot the location of the pinched torus $(M,E)=0$, around 
which $\beta$ has ${4\pi}$ variation. 
Hence, after a parametric loop, we get the mapping ${\vec{\tau}_{a} \mapsto  \vec{\tau}_{a} - \vec{\tau}_{b} }$ 
while ${ \vec{\tau}_{b} }$ remains the same. 
Such non-trivial mapping is the hallmark of monodromy \cite{Duistermaat,monodromybook}.
Upon EBK quantization monodromy in the spectrum is implied, see \App{sE}. 
This is demonstrated in the inset of \Fig{f1}b.  
Namely, transporting an elementary unit cell (spanned by two basis vectors) 
around the pinched torus in the $(M,E)$ spectrum, we end up with a different unit cell.

\rmrk{In \Fig{f1}a the detuning was chosen such that the SP at $n=N/2$ is stable.
Contrary for that, in \Fig{f5} the detuning is such that the SP is at the vicinity of a swap transition. Consequently the spectrum is divided into two regions by the separatrix line, and only the region with the pinched torus exhibits the non-trivial monodromy.} At the swap transition the pinched torus and hence the non-trivial monodromy is relocated to the other region. In the special case of $\mathcal{E}=0$, the pinched torus merges with the separatrix line, leaving both regions with a trivial monodromy.

\section{Summary}

Several themes combine is the study of superflow metastability. There is a monodromy that is associated with the SP that supports the condensate; and a separatrix that is associated with an SP that folds all the depleted states.    
\rmrk{The two SPs determine the skeleton of phase space. By duality it is also the skeleton for the many-body quantum spectrum (via EBK quantization). In the Bloch sphere representation (\Fig{f2}) the two SPs look-alike, but this is in fact wrong and misleading. The topological distinction between the central SP and the peripheral SP becomes conspicuous once we look on the quantum spectrum where the central SP-monodromy appears as a topological defect that reflects the existence of a pinched torus, while the depleted peripheral states form a dense line in ${(M,E)}$ space.}

\rmrk{By itself the above described skeleton is not enough for the understanding of BEC metastability or its absence. The theoretical narrative requires the fusion of {\em chaos} into the story.  If the rotation frequency of the device is adjusted (which controls  the detuning between the central SP and the peripheral separatrix), a stochastic pathway is formed at the ``swap transition", leading to the
depletion of the condensate, and the decay of the superflow.   
In the dual quantum picture chaos blurs the ordered spectrum. Away from the swap transition the topological aspect remains robust , but at the swap transition eigenstates get-mixed and become ergodic within the stochastic region.}

The analysis that we have introduced is specifically relevant for future hysteresis-type experiments \cite{NIST2} with ring lattice circuits \cite{baker,PhysRevLett.119.190403}. \rmrk{Furthermore, the trimer is not only the minimal model for ergodization due to chaos, it is also the minimal configuration for thermalization \cite{trm}, and serves as the building-block for progressive thermalization of large arrays \cite{basko,henning}. Finally, it should be recognized that the theme of metastability is of general interest for mathematical-physics studies of high dimensional chaos, irrespective of specific application.}

\sect{Acknowledgements}
This research was supported by the Israel Science Foundation (Grant No.283/18)
    

%

\appendix


\section{The Trimer Hamiltonian}
\label{sA}

%
For a clean $L$-site ring lattice we define momentum orbitals 
whose wavenumbers are $k=(2\pi/L)\times \text{integer}$. 
Consequently the BHH takes the form 
\be{A2}
\mathcal{H} \ \ = \ \ \sum_{k} \epsilon_k b_k^{\dag}b_k \ + \ \frac{U}{2L} \sum' b_{k_4}^{\dag}b_{k_3}^{\dag}b_{k_2}b_{k_1}
\eeq
where the constraint ${k_1{+}k_2{+}k_3{+}k_4=0}$ mod($2\pi$) is indicate  
by the prime, and ${\epsilon_k =-K \cos (k-\Phi/L)}$ are the single particle energies. 
Later we assume, without loss of generality, that the particles are initially condensed 
in the ${k=0}$ orbital. This is not necessarily the ground-state orbital, 
because we keep $\Phi$ as a free parameter. 
\rmrk{Note that below and in the main text we optionally use~$k$ 
as a dummy index to label the momentum orbitals.}

For the ${L=3}$ site ring we have 
\beq 
\mathcal{H} \ \ &=& 
\ \ \sum_{k=0,1,2} \epsilon_k n_k \ + \ \frac{U}{6} \sum_k n_k^2 \ + \  \frac{U}{3} \sum_{k'\ne k} n_{k'} n_{k} 
\\ \label{eA4} \nonumber
&+& \frac{U}{3} \sum_{k'' \ne k' \ne k}  \left[n_{k'}n_{k''}\right]^{1/2}  \ n_{k}  \ \cos \left( \varphi_{k''} + \varphi_{k'} - 2 \varphi_{k} \right)  
\eeq
We define ${q_1 = \varphi_1 - \varphi_0 }$ and ${q_2 = \varphi_2 - \varphi_0 }$
where the subscripts refers to ${k_{1,2}=\pm(2\pi/3)}$.  
Using the notation ${\mathcal{E}_k=(\epsilon_{k}-\epsilon_{0}) + (1/3)NU}$ we get \Eq{e1} with
\beq \nonumber
\mathcal{H}^{(0)} \ &=& \ \mathcal{E}_1 n_1 + \mathcal{E}_2 n_2 
\ - \ \frac{U}{3} \left[ n_1^2 + n_2^2 + n_1n_2 \right]  
\\ &+& \ \frac{2U}{3} (N{-}n_1{-}n_2) \sqrt{n_1 n_2}  \cos\left(q_1 + q_2\right) 
\eeq
and
\be{A5}
\mathcal{H}^{(+)} \ = \ \frac{2U}{3} \sqrt{(N{-}n_1{-}n_2) n_1} \ n_2 \cos \left(q_1 - 2 q_2 \right)
\eeq
while $\mathcal{H}^{(-)}$ is obtained by swapping the indices~(${1 \leftrightarrow 2}$).  
In fact it is more convenient to use the coordinates  
\be{A6} \nonumber
\phi[\text{mod}(4\pi)] & \ = \ & q_1-q_2  \ =  \  \varphi_1 - \varphi_2 \\
\varphi[\text{mod}(2\pi)] & \ = \ & q_1+q_2 \ =  \ \varphi_1 + \varphi_2 - 2 \varphi_0 
\eeq
and the conjugate coordinates 
\beq
M & \ = \ & \frac{1}{2}(n_1 - n_2) \ \ \in \left[-\frac{N}{2},\frac{N}{2}\right] \\
n & \ = \ & \frac{1}{2}(n_1 + n_2) \ \ \in \left[|M|, \frac{N}{2}\right] 
\eeq
Then the Hamiltonian takes the form of  \Eq{e1} with \Eq{e2} and \Eq{e3},  
where the detuning is ${\mathcal{E} = \mathcal{E}_1+\mathcal{E}_2 - (1/2)NU}$, 
and ${ \mathcal{E}_{\perp} = \mathcal{E}_1 - \mathcal{E}_2 }$.

\section{SPs and seperatrices}
\label{sB}

\hide{The Cherry Hamiltonian is ${ H = \mathcal{E} n + n \sqrt{n+M} \cos(\varphi) }$. 
Below we consider ${H = \mathcal{E} n -  U n^2  + \frac{2U}{3}(N-2n) \sqrt{n^2-M^2} \cos(\varphi) }$.
Later we define the detuning parameter as ${ \varepsilon = 6(\mathcal{E}/NU) - 3 }$. 
With this convention the energy along the ${n=N/2}$ perimeter is ${H = (\varepsilon/12)N^2U}$}

Consider a phase-space that is described by ${\varphi,n}$. We shall distinguish between {\em rotor geometry} for which the ${n=0}$ points are distinct, and {\em oscillator geometry} for which all the  ${n=0}$ points are identified as one point. The algebraic treatment is the same, but the physical interpretation is different.

\sect{Regular point}
As an appetizer consider  
the Hamiltonian 
\be{B0}
H \ \ = \ \ \sqrt{2n} \sin(\varphi)
\eeq
It looks singular at ${n=0}$, but in fact it is completely smooth there. Regarded as an oscillator it is canonically equivalent to ${ H = p }$ that generates translations. Similar observation applies to the non-interacting dimer Hamiltonian ${ H=(1/2)(a_2^{\dag}a_1 + \text{h.c.}) }$, which in action angle variables takes the form   
\beq
H \ \ = \ \ \sqrt{\left(\frac{N}{2}-n\right)\left(\frac{N}{2}+n\right) } \ \cos(\varphi)
\eeq
Here both the North and the South  poles of the Bloch sphere (${n=\pm N/2}$) are regular phase-space points, neither SP nor singular.

\sect{Stationary point}
Consider the standard quadratic Hamiltonian ${H=(1/2)[a p^2 + b x^2]}$. In polar canonical coordinates it is 
\be{B1}
H \ \ = \ \ \left[A+B\cos(2\varphi) \right] \ n 
\eeq
with ${A=(a+b)}$ and ${B=(a-b)}$. If ${ab>0}$, equivalently ${|A|>|B|}$, the origin (${n=0}$) is an elliptic SP that is circled by trajectories that have the frequency 
\be{B2}
\omega \ = \ \sqrt{ab}  \ = \ \sqrt{A^2-B^2} 
\eeq
Otherwise the origin becomes an unstable hyperbolic SP. In the latter case there is an 8-like separatrix that goes through the origin: there are two ingoing directions and two outgoing directions. The approach to the SP along the separatrix, and its departure, is an infinitely slow motion. 
\rmrk{The SPs of the dimer \Eq{eA3} are described locally by the above Hamiltonian.}

\sect{Folded SP}
Consider the dimer Hamiltonian \Eq{eA3} with $2\tilde{\varphi}$ replaced by $\varphi$. Here the dynamics is the same from algebraic perspective, but the global geometry is different. It is a folded version of the dimer Hamiltonian. In the hyperbolic case the vicinity of the SP can be described as ``half saddle". From local dynamics point of view the equations of motion are identical, but here the separatrix has only one outgoing direction and only one ingoing direction. 
%

\sect{Spread SP}
Consider \Eq{eB1}, but assume that we are dealing with rotor geometry. From local dynamics point of view the equations of motion are still identical, but now the arrival point (say ${ \varphi_{in} }$) and the departure point (say ${ \varphi_{out} }$) are not the same point.

\sect{Stability analysis}
Consider the Hamiltonian of \Eq{e2} with ${M=0}$. 
The origin (${n=0}$) is a {\em folded SP}. 
It is elliptic or hyperfolic  depending on the detuning. 
Locally the Hamiltonian looks like \Eq{eB1} with 
\beq
&& A = \mathcal{E} +\frac{NU}{2}, \ \ \ \ \ B = \frac{2NU}{3}  \\ \label{eB4}
&& \text{SP unstable for} \ \ -\frac{7NU}{6} <  \mathcal{E} <  \frac{NU}{6}
\eeq
In the regime where the SP is stable the $\omega$ of \Eq{eB2} 
reflects the frequency of the Bogoliubov excitations \cite{sfc}. 
In the hyperbolic case we have a separatrix that goes through the origin.

For the same Hamiltonian, the perimeter (${n=N/2}$) is a {\em spread SP}.
For the purpose of stability analysis we can identify the points along the perimeter 
as a single point of a Bloch sphere. Setting $\tilde{n}=N-2n$ the Hamiltonian 
looks like \Eq{eB1} with 
\beq
&& A = -\frac{\mathcal{E}}{2} +\frac{NU}{4}, \ \ \ \ \ B = \frac{NU}{3}  \\ \label{eB5}
&& \text{SP unstable for} \ \ -\frac{NU}{6} <  \mathcal{E} <  \frac{7NU}{6}
\eeq
In the hyperbolic case we have a separatrix that meets the perimeter at one point and departs in a different point. Combining with \Eq{eB4} we see that both SPs are unstable if ${|\mathcal{E}| < (1/6)NU}$.
In the latter case we have two seperatrices. The separatrices swap as we go through ${\mathcal{E}=0}$, see \Fig{f1}.

\sect{The case of nonzero $M$}
For the same Hamiltonian \Eq{e2} with ${M \ne 0}$, 
the points along the inner boundary ${n=M}$ are distinct. 
So we cannot regard them as a single point. 
Close to this inner boundary we have ${H \sim \sqrt{\tilde{n}}\cos(\varphi) }$, 
with ${\tilde{n}=n-M}$.  This is a non-singular Hamiltonian, 
essentially the same as \Eq{eB0}, that generates regular flow. 
It follows that the inner boundary is not special from a dynamical point of view: 
it can be regarded as {\em spread regular point}, 
it is not an SP, and there is no separatrix there. 

The stability of the perimeter is determined as in \Eq{eB5}, but with $B$ multiplied by $\sqrt{1-(2M/N)}$. Therefore, for sufficiently large $M$ we always have  ${|A|>|B|}$ and the perimeter is stable.

\begin{figure}
\begin{center}
\begin{overpic}[width=0.3\hsize]{cone1} \put (22,95) {(a)} \end{overpic}
\begin{overpic}[width=0.3\hsize]{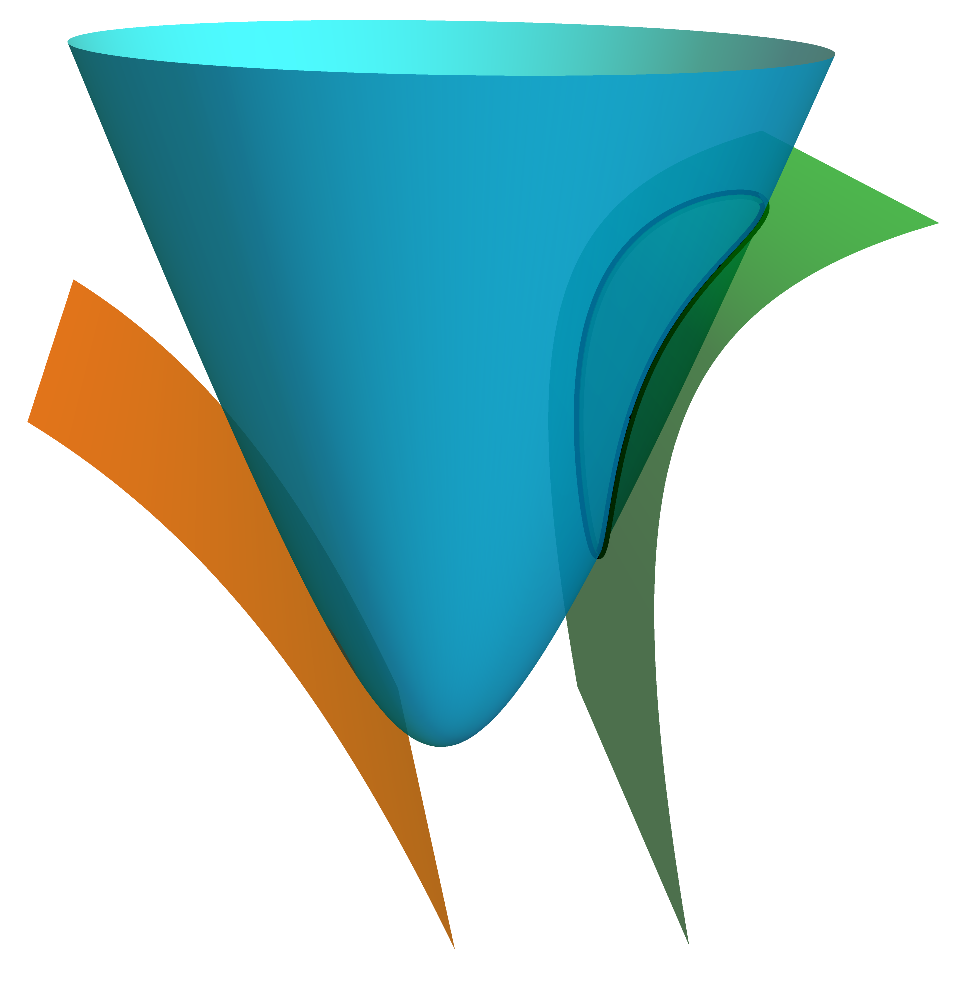} \put (22,99) {(b)} \end{overpic}
\begin{overpic}[width=0.3\hsize]{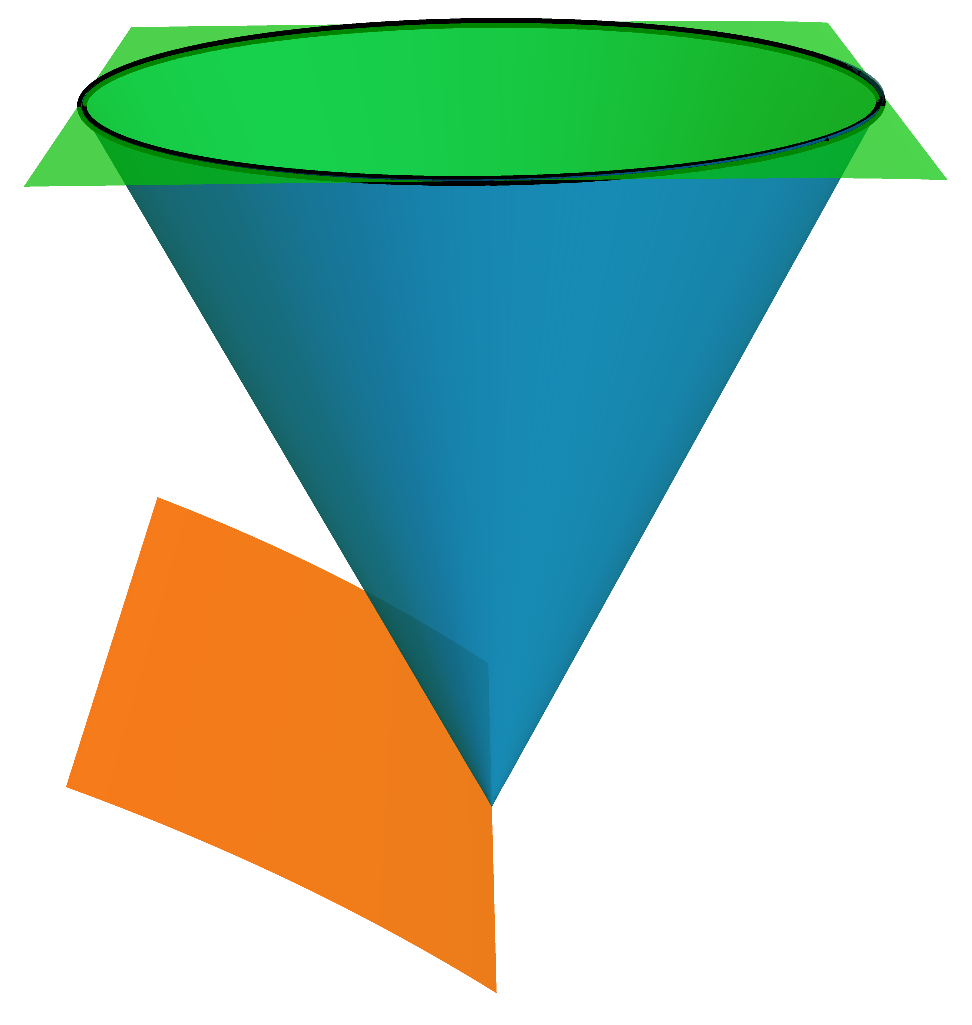} \put (22,99) {(c)} \end{overpic}
\begin{overpic}[width=0.3\hsize]{cone4} \put (22,95) {(d)} \end{overpic}
\begin{overpic}[width=0.3\hsize]{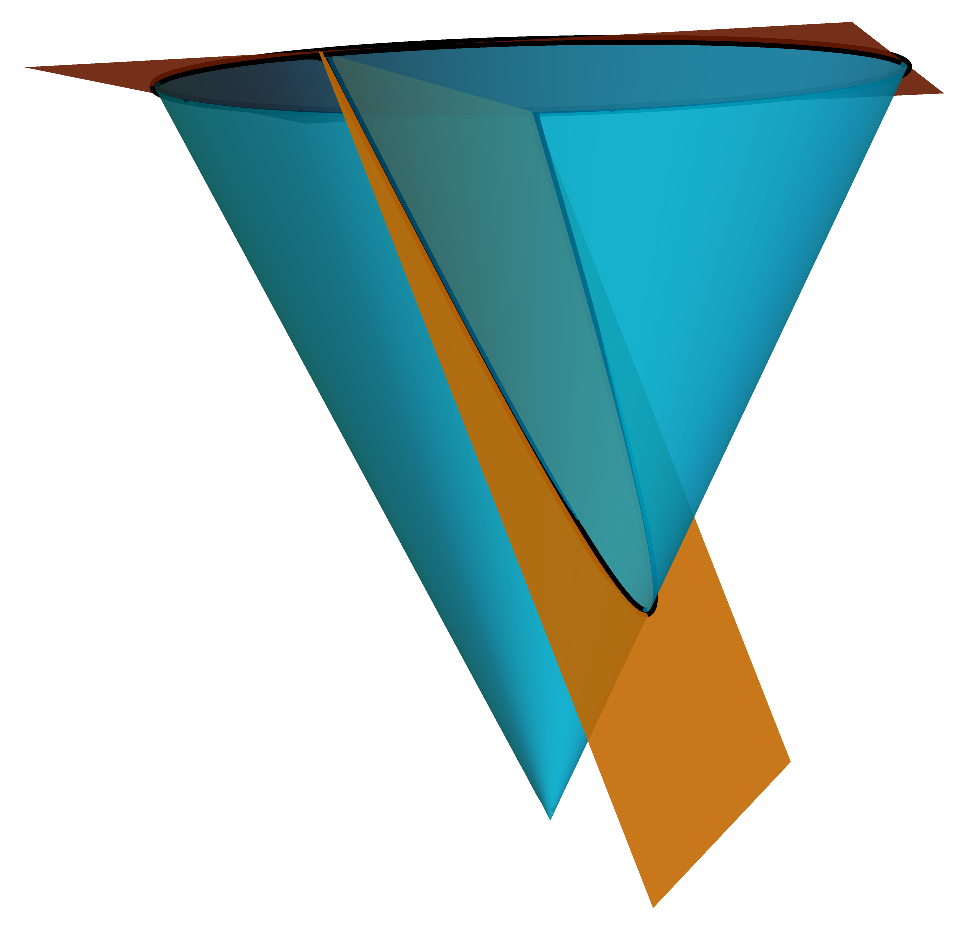} \put (22,95) {(e)} \end{overpic}
\begin{overpic}[width=0.3\hsize]{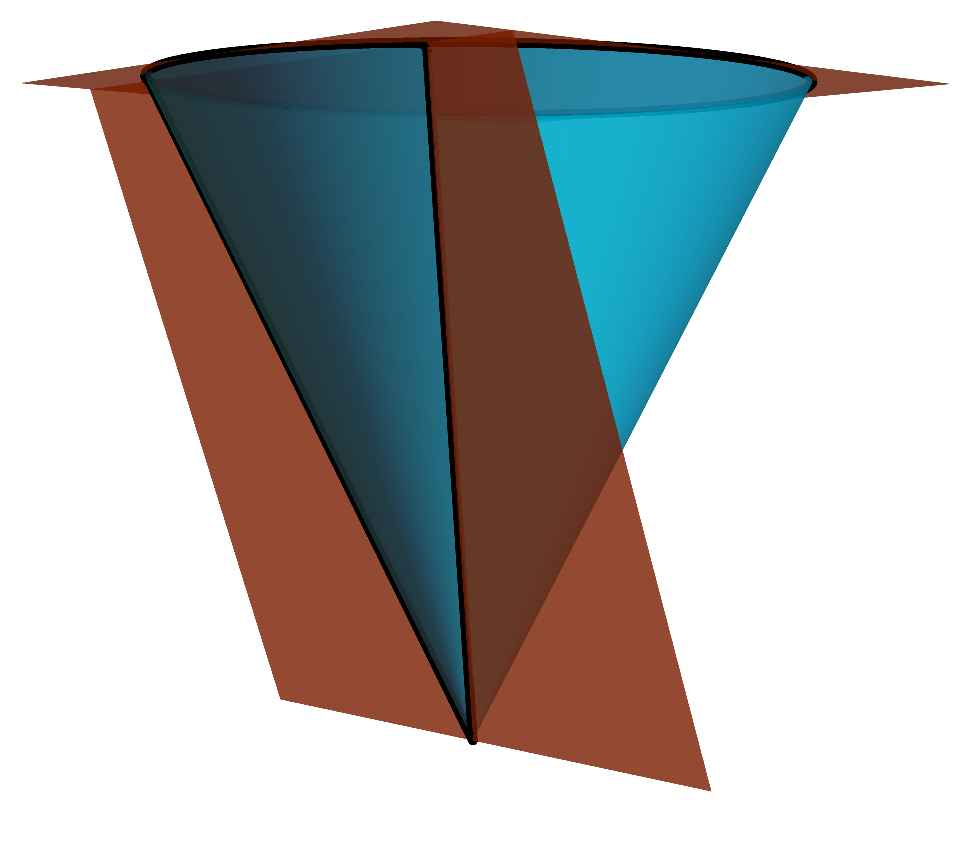} \put (22,90) {(f)} \end{overpic}
\caption{ \label{fs4}  Several examples of the reduced phase-space in the $(K_x,K_y,K_z)$ coordinates (the symmetry axis is $K_z$). 
The blue cone (a,c-f) is the surface of constant $M=0$ while the blue hyperbole (b) is the surface of constant $M=0.15N$. The remaining surfaces correspond to a constant $E$. 
The intersection of constant $M$ and $E$ surfaces (highlighted in black) is a trajectory in the reduced $(\varphi,n)$ space,
and provides a useful way of visualizing the phase space tori (see text).}
\end{center}
\end{figure}

\begin{figure*}
\begin{center}
\includegraphics[width=0.13\hsize]{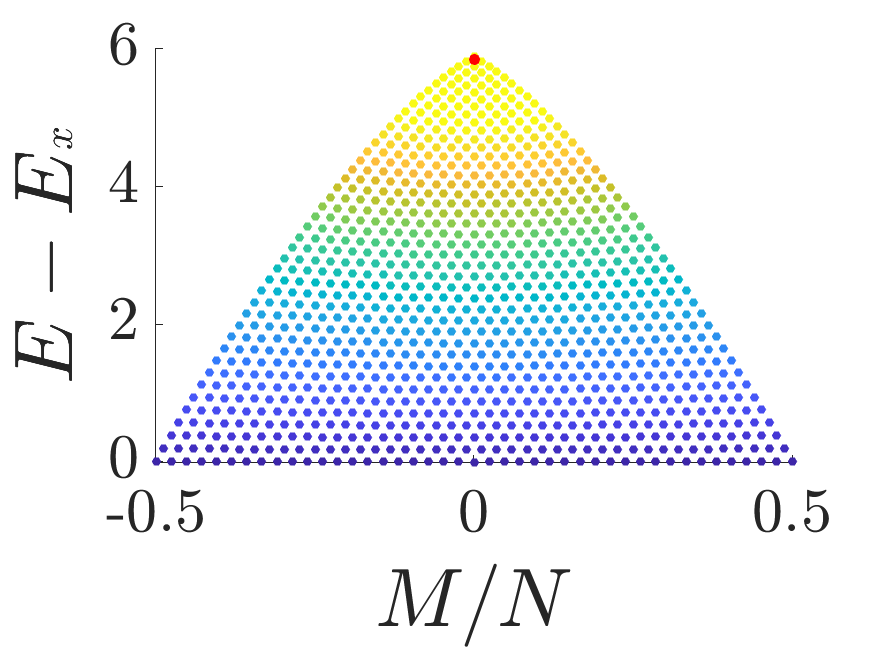} 
\includegraphics[width=0.13\hsize]{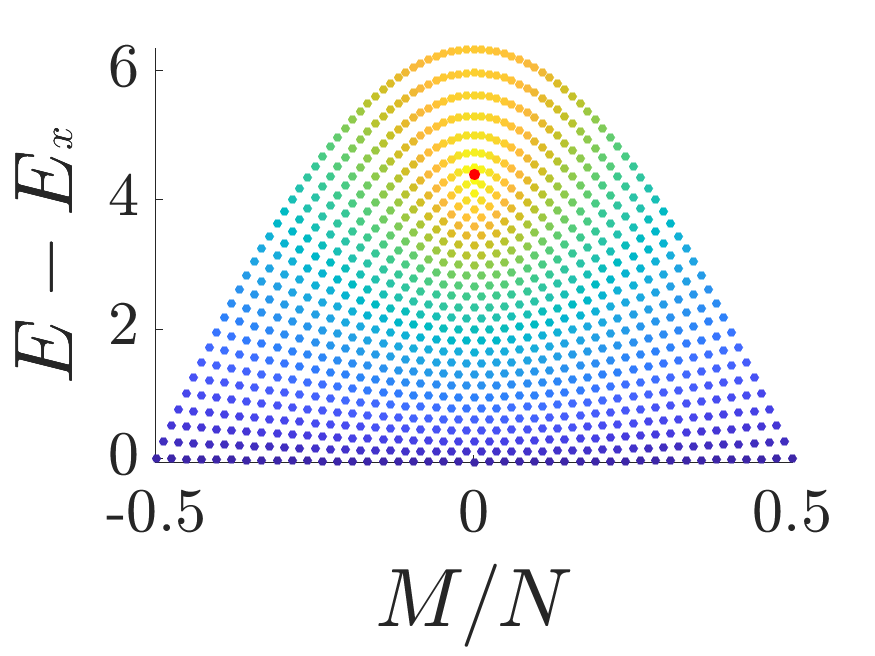} 
\includegraphics[width=0.13\hsize]{specm03eta100} 
\includegraphics[width=0.13\hsize]{spec0eta100} 
\includegraphics[width=0.13\hsize]{spec03eta100} 
\includegraphics[width=0.13\hsize]{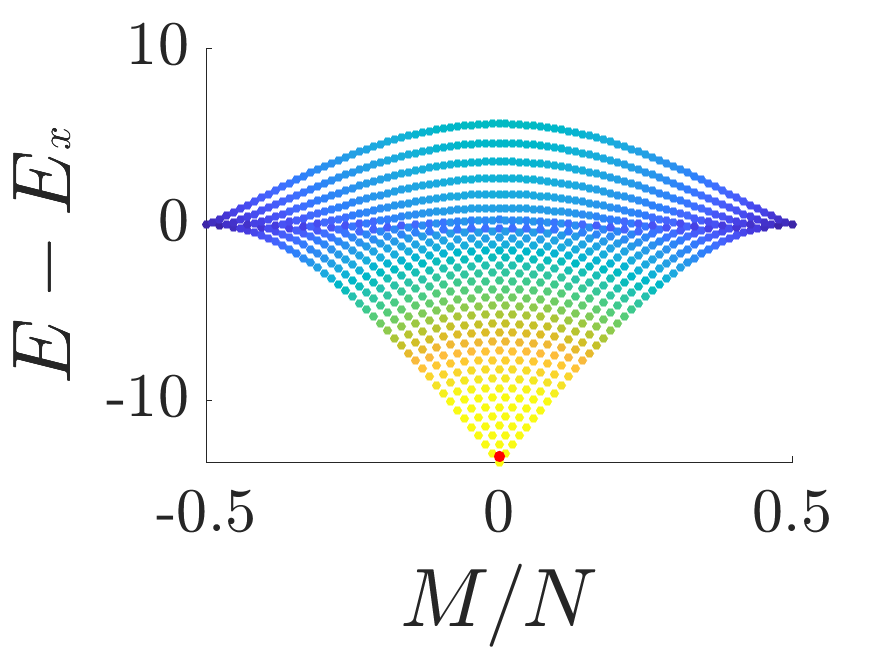} 
\includegraphics[width=0.13\hsize]{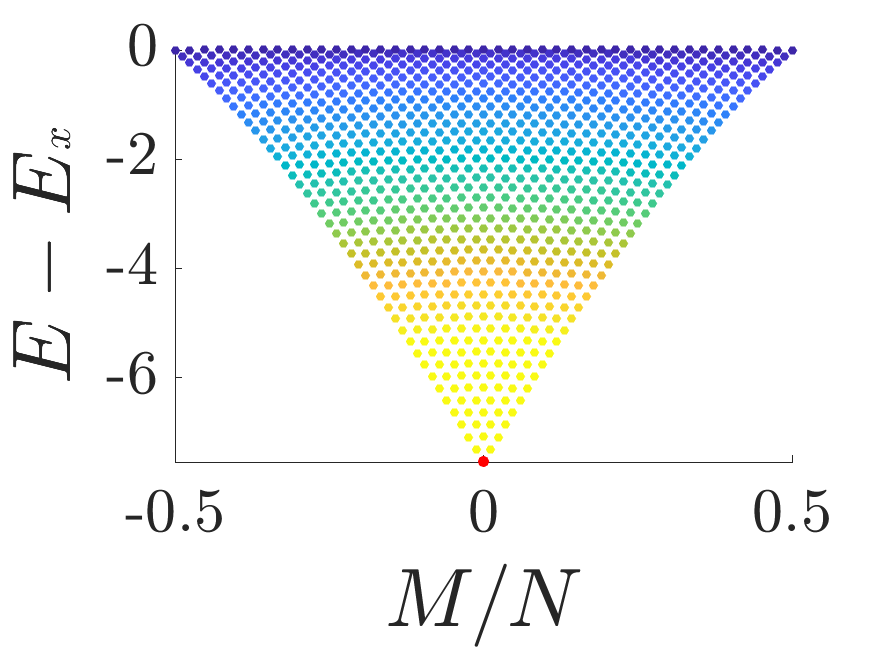} 
\includegraphics[width=0.13\hsize]{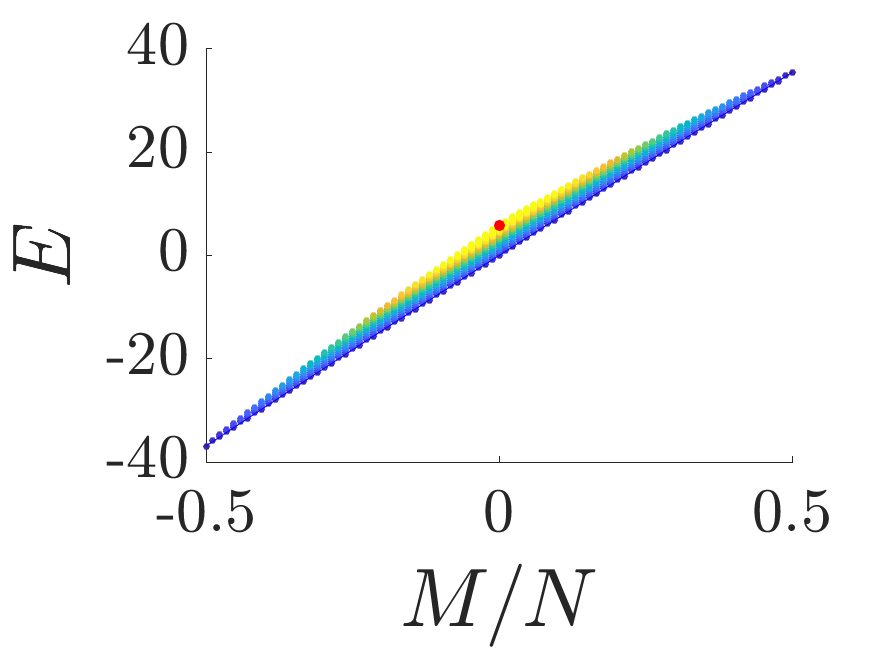} 
\includegraphics[width=0.13\hsize]{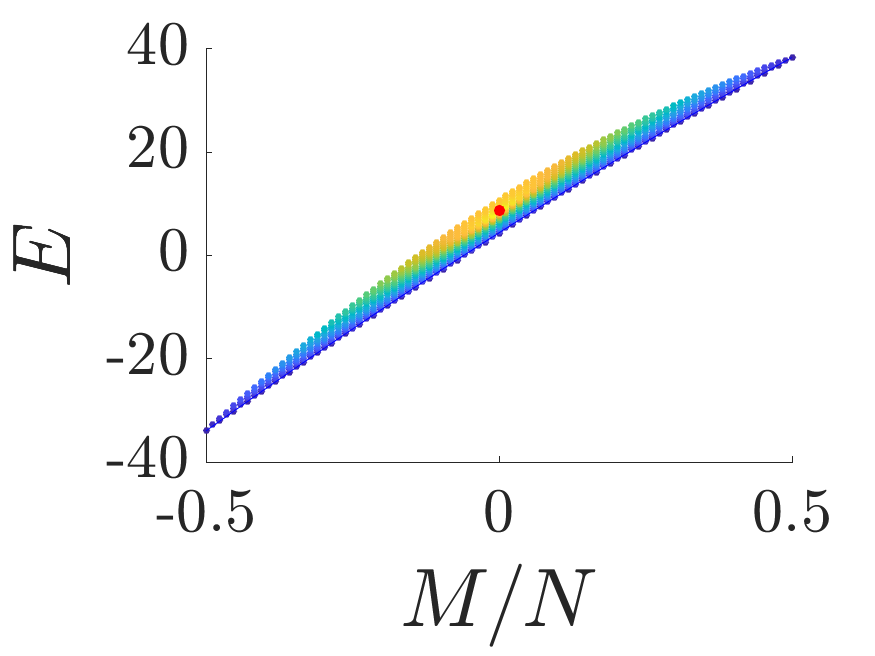} 
\includegraphics[width=0.13\hsize]{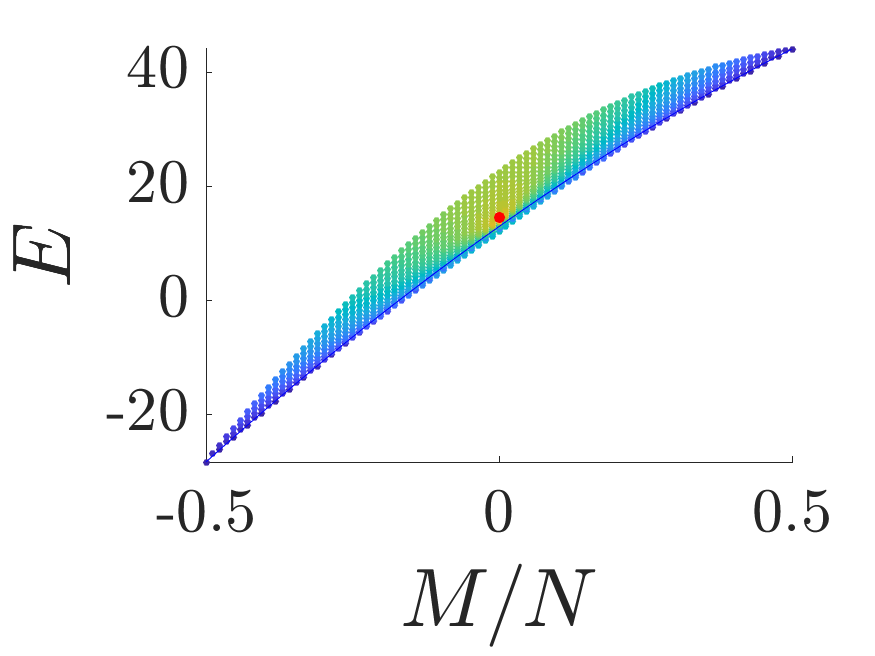} 
\includegraphics[width=0.13\hsize]{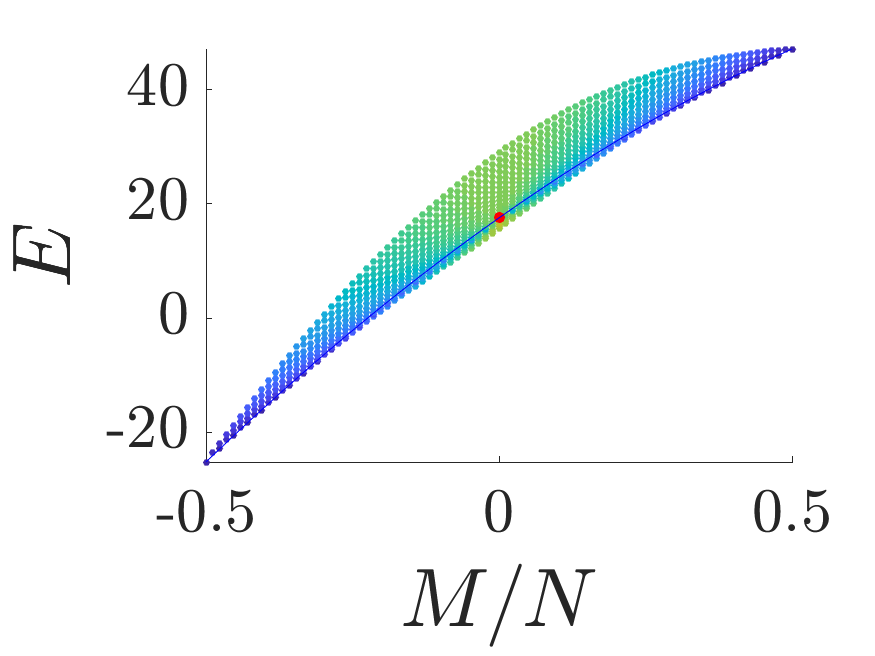} 
\includegraphics[width=0.13\hsize]{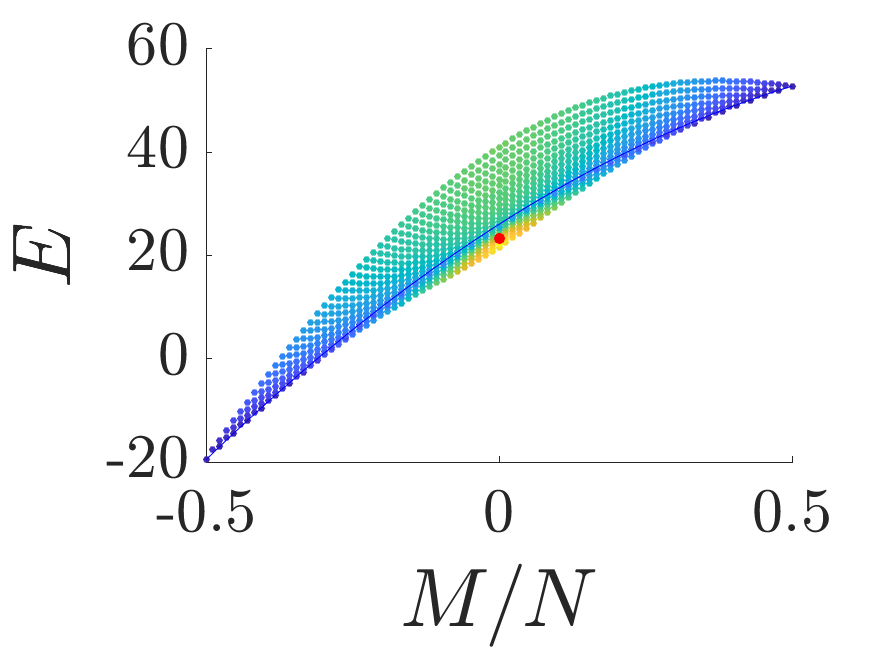} 
\includegraphics[width=0.13\hsize]{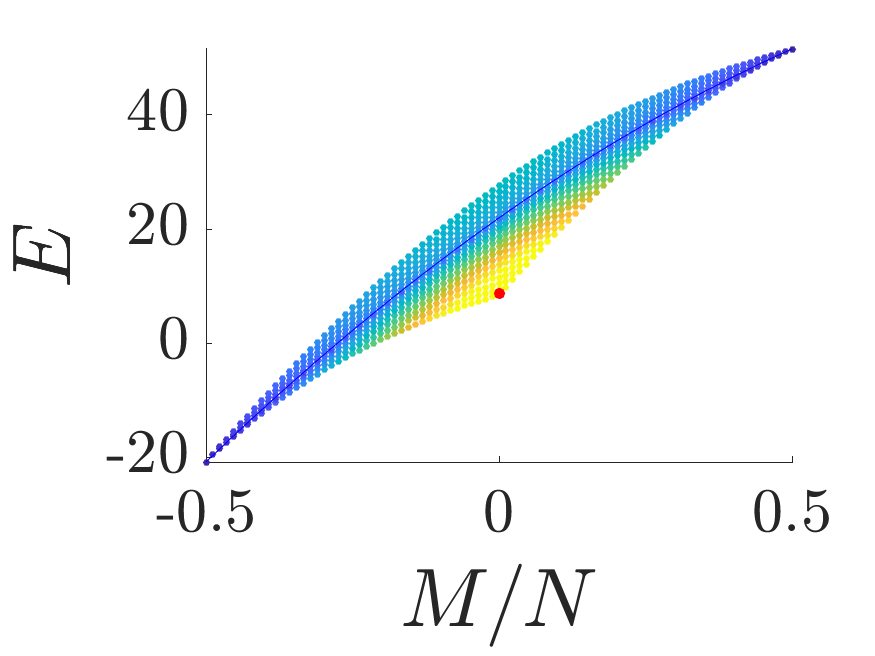} 
\includegraphics[width=0.13\hsize]{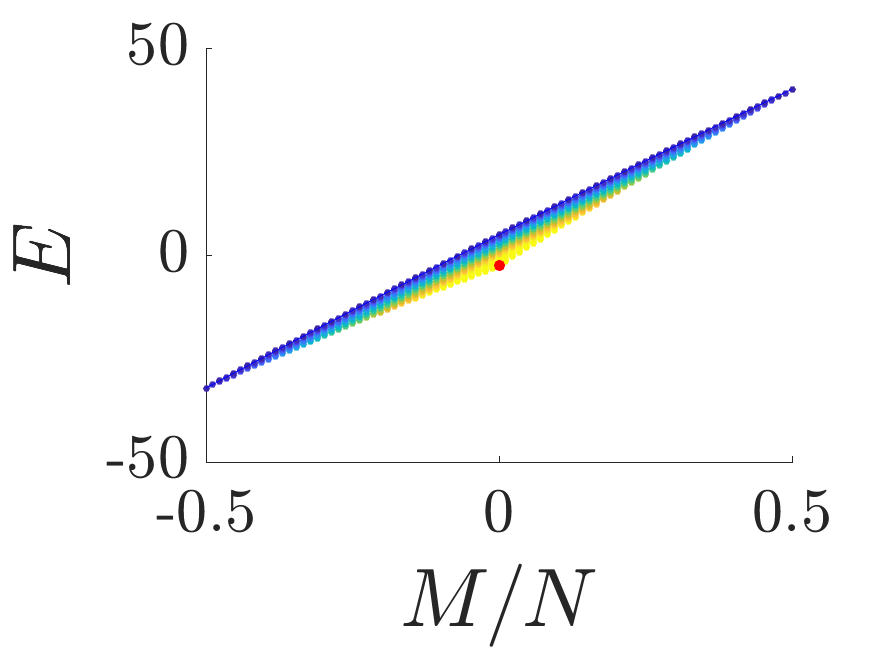} 
\caption{ \label{fs3} 
{\bf The spectrum.} 
The top row panels are the spectrum of $\mathcal{H}^{(0)}$ with the same $\mathcal{E}/NU$ values as in \Fig{f2}. 
In the bottom row the same spectra is plotted, but without subtracting the separatrix energy. The interaction strength from left to right is $NU \approx 0.2 , 0.6,1.4,1.9 , 2.7 , 1.9 , 0.3 $ in units of the BHH hopping frequency $K$. Note that for the top row panels, different $NU$ values will produce the same plot and only scale the $E-E_x$ axis. Note that the energy here differs by a constant from \Fig{f1}(b).
\hfill
} 
\end{center}
\end{figure*}

\section{Conical intersection perspective}
\label{sC}

A useful way for visualizing the phase space tori is based on the $SU(1,1)$ symmetry \cite{SU11,PhysRevA.83.033605} of $\mathcal{H}^{(0)}$.
For that we express the two conserved quantities, namely the energy $E$ and the constant of motion $M$, in terms of the group generators. 
We start by introducing:
\beq
K_z=n+ \frac{1}{2} \ \ \ , \ \ \    K_+ = a_1^\dagger a_2^\dagger   \ \ \ , \ \ \     K_- = a_1 a_2 
\eeq
which is a realization of the $SU(1,1)$ group, satisfying the algebra:
\beq
[K_z,K_\pm] =\pm K_\pm  \ \ \ , \ \ \ [K_+,K_-] =-2 K_z
\eeq
The Casimir operator of the group, which commutes with all the generators, is:
\beq
C \ = \ K_z^2 - K_x^2 - K_y^2 \label{eC1}
\eeq
where $K_x$ and $K_y$ are given by $ K_\pm = K_x \pm i K_y  $.
In the semiclassical treatment we have:
\beq
K_x & \ \sim \ &  \sqrt{n_1 n_2}\cos \varphi  \ \ \in \left[- \Delta, \Delta  \right] \\
K_y & \ \sim \ &  \sqrt{n_1 n_2}\sin  \varphi  \ \ \in \left[- \Delta, \Delta  \right] \\
K_z & \ \sim \ & n    \ \ \in \left[|M|, \frac{N}{2}\right] 
\eeq
where $\Delta= \sqrt{(N/2)^2 -M^2 }$.
The Hamiltonian can be written in terms of the generators as:
\beq
\mathcal{H}^{(0)}  \ & = & \ \mathcal{E} K_z  + \mathcal{E}_{\perp} M - \frac{U}{3}M^2   \\ \nonumber
&+& \ \frac{2U}{3} (N-2K_z)\left[ \frac{3}{4}K_z + K_x \right] 
\eeq  
As for the constant of motion $M$, we have $M^2=C$.
In \Fig{fs4} we plot several examples for the $M^2$ and $ \mathcal{H}^{(0)} =E$ surfaces in the $(K_x,K_y,K_z)$ space. 
For $ M =0 $  \Eq{eC1} defines a cone whose tip corresponds to ${n=0}$, while its outer boundary to ${n=N/2}$.
For for a constant $M \neq 0$  \Eq{eC1} defines an hyperboloid whose base corresponds to ${n=|M|}$, while its outer boundary to ${n=N/2}$.

The intersection between the $E$ and $M^2$ surfaces is a trajectory in the reduced $(\varphi,n)$ phase space.
In the full phase space, we also have the phase $\phi$, which dynamically changes as ${\dot{\phi} =  \partial \mathcal{H}^{(0)} / \partial M}$. If the intersection between the surfaces forms a closed loop, as in \Fig{fs4}(a,b), the dynamics in the full $(\varphi,\phi,n,M)$ phase-space covers a 2-torus (which is, of course, the typical case in an integrable 2 DOF system).
When the two surfaces tangent, as in \Fig{fs4}(b), the trajectory is a fixed point in the $(\varphi,n)$ space, and a circle in the full phase space.

Trajectories that pass through $n=0$ should be addressed with more caution.
As explained in the main text, the tip of the cone does not correspond to a $\phi$ circle, but to a single point. This is because $n=0$ means $n_1=n_2=0$ so that $\phi$ is degenerate. When the $n=0$ SP is stable, the energy surface is tangent to the tip of the cone, as in \Fig{fs4}(c), hence the trajectory is a single point in phase space.
When unstable, the intersection forms a cusped circle, see \Fig{fs4}(d), representing a {\em pinched torus}, i.e. a torus with one of its $\phi$ circles shrinks to a point.

Trajectories that pass through $n=N/2$ are special too.
When the $n=N/2$ SP is stable, as in \Fig{fs4}(c), the intersection is the entire outer circle of the cone, reflecting the fact that it is a {\em spread} SP. When unstable, see \Fig{fs4}(e), a separatrix trajectory is formed, which meets the $n=N/2$ circle at two points, corresponding to ${ \varphi_{\text{in}} }$ and ${ \varphi_{\text{out}} }$.
At the swap transition, the two SPs are connected, i.e. the cusped circle of $n=0$ merged with the separatrix trajectory of $n=N/2$, as shown in \Fig{fs4}(f).

\section{Gallery}
\label{sD}

Here we provide additional plots of the spectra for the whole range of the detuning parameter.
\Fig{fs3} is an extended version of \Fig{f5} and corresponds to \Fig{f2}.

\section{Hamiltonian Monondromy}
\label{sE}

Consider generators ${ (H_1, H_2) }$ in involution, i.e. that commute with each other. 
The generated trajectories are moving on an energy surface labeled ${ (E_1, E_2) }$.
A {\em walk} consists of ${t_1}$ evolution using ${H_1}$, and ${t_2}$ evolution using ${H_2}$. 
The involution implies that the walks are commutative. 
Accordingly the parameterization of a walk is ${ \vec{t} = (t_1, t_2) }$.
Periodic walk is a walk that brings you back to the same point.
The set of periodic walks forms a lattice in ${\vec{t}}$ space. 
This lattice is spanned by basis vectors ${\vec{\tau}_k}$, where ${k=a,b}$.
We can formally write any point in  ${\vec{t}}$ space as  
\beq 
\vec{t} \ \ = \ \ \sum_k \frac{\theta_k}{2\pi} \vec{\tau}_k \ \ = \ \ \frac{\theta_{a}}{2\pi} \vec{\tau}_{a}  +  \frac{\theta_{b}}{2\pi} \vec{\tau}_{b}  
\eeq 
We define a reciprocal basis such that 
\beq
\vec{\omega}_k \cdot \vec{\tau}_{k'} \ \ = \ \ 2\pi \delta_{k,k'} 
\eeq
The reciprocal relation is 
\beq
\theta_k \ \ = \ \ \vec{\omega}_k \cdot \vec{t} 
\eeq 
Once action variables are defined we have 
\beq
\vec{\omega}_k \ \ =  \ \ \left( \frac{\partial H_1}{\partial J_k},  \frac{\partial H_2}{\partial J_k} \right)  
\eeq
The spacings between two energies is 
\beq
\Delta E \ \ = \ \ \sum_k \vec{\omega}_k \cdot \ora{\Delta n}_k 
\eeq
Thus the spectrum forms a reciprocal lattice.

Considering a closed loop in ${ (E_1, E_2) }$ space, the monodromy matrix is defined as the mapping
\beq
\vec{\tau}_k(\text{final}) \ \ &=& \ \ \sum_l   \mathbf{M}_{kl}    \vec{\tau}_l 
\eeq
If the loop encircles a pinched torus we have \cite{monodromybook}
\beq 
 \mathbf{M}   \ = \  
\left( 
\begin{array}{cccc}
1 & -1 \\
0 & 1 \\
\end{array}
\right) 
\eeq
so we get the mapping ${\vec{\tau}_{a} \mapsto  \vec{\tau}_{a} - \vec{\tau}_{b} }$, 
as discussed in the main text after \Eq{e4}. 
For the reciprocal basis we have:
\beq
\vec{\omega}_k(\text{final}) \ \ &=& \ \  \sum_l   \tilde{\mathbf{M}}_{kl} \  \vec{\omega}_l 
\eeq
where ${ \tilde{\mathbf{M}} = [\mathbf{M}^{-1}]^t }$. This can be seen by writing:
\beq
&& 2 \pi \delta_{k,k'} \ = \ \  [ \vec{\omega}_k(\text{final}) ]  \cdot   [ \vec{\tau}_{k'}(\text{final})] 
\\ \nonumber  
&& \ \ \ = \ \sum_{lm}  \tilde{\mathbf{M}}_{kl}  \mathbf{M}_{k'm} \ \vec{\omega}_i  \cdot \vec{\tau}_j  
\ = \ 2 \pi \sum_{l}  \tilde{\mathbf{M}}_{kl}  \mathbf{M}_{k'l} 
\eeq
hence $   \tilde{\mathbf{M}}   \mathbf{M}^t  = \mathbf{1}    $ and ${ \tilde{\mathbf{M}} = [\mathbf{M}^{-1}]^t }$.
For a loop which encircles the pinched torus we then have
\beq 
 \tilde{\mathbf{M}}  \ = \  
\left( 
\begin{array}{cccc}
1 & 0 \\
1 & 1 \\
\end{array}
\right) 
\eeq
which reflects the way $\vec{\omega}_k$ are mapped, and therefore how a unit cell in the quantum spectrum is transformed, as seen in \Fig{f1}(b).

\clearpage
\end{document}